\documentclass[amsmath,amssymb,twocolumn]{revtex4-1}

\usepackage{graphicx,mathtools}
\usepackage[varg]{txfonts}
\usepackage{array}
\usepackage{simpler-wick}
\usepackage{color}
\usepackage{float}
\usepackage[british]{babel}
\usepackage[T1]{fontenc}
\usepackage[utf8]{inputenc}

\newcommand{\ii}{\mathrm{i}}
\newcommand{\e}{\mathrm{e}}
\renewcommand{\d}{\mathrm{d}}

\newcommand{\tens}[1]{\mathbf{#1}}

\newcommand{\nexp}[1]{\mathrm{exp}\left\{ #1 \right\}}
\renewcommand{\o}[0]{\,\hat\Phi}

\newcommand{\ntop}[0]{{\, \top}}
\newcommand{\ave}[1]{\left\langle #1 \right\rangle}
\DeclareMathOperator{\dirac}{\delta_\textsc{d}}

\newcommand{\der}[2]{\frac{\delta{#1}}{\ii \delta{#2}}}

\renewcommand{\lim}[2]{\,\underset{\vec q_{#1} \rightarrow \vec q_{#2}}{\mathrm{lim}}\,}

\begin{document}

\title{A Model for Hydrodynamics in Kinetic Field Theory}
\author{C.\ Viermann\textsuperscript{1,2}, J.T.\ Schneider\textsuperscript{1,3}, R.\ Lilow\textsuperscript{1,4}, F.\ Fabis\textsuperscript{1}, C.\ Littek\textsuperscript{1}, E.\ Kozlikin\textsuperscript{1}, M.\ Bartelmann\textsuperscript{1}}
\affiliation{\textsuperscript{1} Heidelberg University, Zentrum f\"ur Astronomie, Institut f\"ur Theoretische Astrophysik, Philosophenweg 12, 69120 Heidelberg, Germany}
\affiliation{\textsuperscript{2} Heidelberg University, Kirchhoff-Institut f\"ur Physik, Im Neuenheimer Feld 227, 69120 Heidelberg}
\affiliation{\textsuperscript{3} Heidelberg University, Institut f\"ur Theoretische Physik, Philosophenweg 16, 69120 Heidelberg}
\affiliation{\textsuperscript{4} Department of Physics, Technion, Haifa 3200003, Israel}

\begin{abstract}

	In this work, we introduce an effective model for both ideal and viscous fluid dynamics within the framework of kinetic field theory (KFT). The main application we have in mind is cosmic structure formation where gaseous components need to be gravitationally coupled to dark matter. However, we expect that the fluid model is much more widely applicable. 

	The idea behind the effective model is similar to that of smoothed particle hydrodynamics. By introducing mesoscopic particles equipped with a position, a momentum, and an enthalpy, we construct a free theory for such particles and derive suitable interaction operators. We then show that the model indeed leads to the correct macroscopic evolution equations, namely the continuity, Euler, Navier-Stokes, and energy conservation equations of both ideal and viscous hydrodynamics.

\end{abstract}

\maketitle

\section{Introduction}

	In upcoming years, powerful weak gravitational lensing studies will hopefully deliver detailed data on the matter distribution in the universe. To gain meaningful insight, these measurements must be combined with accurate predictions of the cosmological power spectrum and higher order spectra for all matter.
	
	In past works, we presented an analytic approach to predicting the dark matter power spectrum. This approach is based on a kinetic field theory for classical particles (KFT) originally developed by Das and Mazenko (\cite{Mazenko1, Mazenko2, Mazenko3}) to describe the ergodic to non-ergodic transition in glasses. It was then adapted to cosmic structure formation by Bartelmann et. al (\cite{2016NJPh...18d3020B}).
	
	KFT allows the prediction of correlation functions of macroscopic quantities (like density or velocity-density) based only on initial correlations and microscopic particle dynamics. First very promising results were achieved that closely match simulations (\cite{2018JSMTE..04.3214F, 2017NJPh...19h3001B, 2016NJPh...18d3020B, 2015PhRvE..91f2120V, 2015PhRvD..91h3524B, 2399-6528-2-2-025020}).
	
	So far, calculations within KFT are limited to ensembles of weakly interacting particles like dark matter particles whose trajectories are only slightly perturbed (which can nonetheless lead to large density perturbations).
	
	For many cosmological applications, the dynamics of dark matter must be coupled to the dynamics of `ordinary' matter. The gaseous ordinary matter behaves fluid-like, i.e.\ it is in a dynamical regime where particles undergo frequent and strong interactions. As we will argue in more detail below, within KFT this strongly interacting regime is difficult to reach from fundamental principles, although it must be conceptually possible.
	
	As an alternative, we present an effective model for fluid ensembles in the KFT description. The basic idea of this model is similar to the conventional derivation of hydrodynamics and particle-based schemes in numerical fluid codes like smoothed particle hydrodynamics (SPH).
	
	While our main motivation is the cosmological matter power spectrum for mixtures of dark and fluid-like matter, the fluid model in KFT is not limited to this application. It might (in a resummed form) give access to statistical properties of different fluid phenomena, maybe even to turbulent spectra. A combination of the hydrodynamical model with electromagnetic particle interactions might allow the study of magnetohydrodynamics and related instabilities.
	
	Within this work, we introduce the model of hydrodynamics in the framework of KFT, show that the corresponding macroscopic evolution equations match the continuity equation, the Euler or Navier-Stokes equation, respectively, and the energy conservation equation of hydrodynamics. Finally, we present an exemplary calculation of how a localised high-pressure region evolves dynamically within the KFT fluid model.
	
	Before we shift our focus to hydrodynamics, we briefly introduce KFT for readers unfamiliar with the formalism.

\section{Brief summary of KFT} \label{summary_KFT}
	We follow the notation introduced in \cite{2016NJPh...18d3020B}, to which we refer readers for a more thorough derivation. Here, we only provide a brief summary which is partly taken from \cite{2015PhRvE..91f2120V}. The latter work also contains an exemplary calculation within KFT which might help in understanding the formalism.

	The derivation of this non-equilibrium statistical field theory begins with the equation of motion of an $n$-component classical field $\varphi_a$ in $d$ space-time dimensions
	\begin{align} \label{general_equation_of_motion}
		E(\varphi_a) =   \dot {\varphi}_a + E_0(\varphi_a) + E_\mathrm I (\varphi_a) = 0\,,
	\end{align}
	where all terms describing free motion are bundled into $E_0$ and all interaction terms between particles or with external fields are part of $E_\mathrm I$.

	For a classical theory, the field can only evolve from an initial state into a final state if the transition satisfies the above equation of motion. Starting from this requirement and setting the interaction part of the equation of motion to zero, for the moment, one can derive the centre piece of the theory, i.e.\ its free generating functional
	\begin{align}\label{general_free_functional}
		Z_0[J, K] =& \int \d\Gamma_{\mathrm i} \,\nexp{\ii \int_{\mathrm i}^{\mathrm f} \d t \, J_a \bar \varphi_a }\;,
	\end{align}
	where $J_a$ was introduced as a source field conjugate to the degrees of freedom $\varphi_a$. The free trajectory of each degree of freedom is denoted by an overbar, $\bar \varphi_a$, and is defined as
	\begin{equation}\label{general_phi_bar}
		\bar \varphi_a(t) = G_\mathrm R (t,t_\mathrm i)\, \varphi_a^{\,\mathrm{(i)}} - \int_{t_\mathrm{i}}^t\d t'\,G_\mathrm R(t,t')\, K_a(t') \;,
	\end{equation}
	where $\varphi_a^{\,\mathrm{(i)}}$ is the initial field configuration, $G_R$ the retarded Green's function of the free equation of motion and  $K_a$ a second source field that will later allow to introduce interactions. More precisely, it is the source field to an auxiliary field $\chi_a$ that needs to be introduced in the derivation of the generating functional. In the free functional, the  $\chi_a$ are already integrated out but will reappear once interactions are switched on. Finally, $\mathrm d \Gamma_{\mathrm i}$ denotes the integration over the initial state space for all degrees of freedom weighted by an initial probability distribution.

	So far, the interacting part of the equation of motion has been neglected. If it is included, the generating functional can be brought into the form
	\begin{align}\label{general_interacting_functional}
		Z[J,  K] =& \e^{\ii\hat S_\mathrm I }  \int \d\Gamma_{\mathrm i}\, \nexp{\ii \int_{\mathrm i}^{\mathrm f} \d t \,  J_a\bar\varphi_a} \;.
	\end{align}
	Here we introduced the interaction term $S_\mathrm I$  whose operator representation $\hat S_\mathrm I$ is defined by
	\begin{align}\label{general_interaction_operator}
		S_\mathrm I = \int \d t \, E_\mathrm I (\varphi_a) \, \chi_a \quad \leftrightarrow \quad \hat S_\mathrm I = \int \d t\,  E_\mathrm I \left(\der{}{J_a(t)}\right)\,  \der{}{K_a(t)}\;,
	\end{align}
	with an implied summation over $a$. In Eq.~\eqref{general_interacting_functional}, the $K_a$-derivative contained in the interaction operator acts on the inhomogeneous term in $\bar \varphi_a$ and thus alters the time evolution of the respective degree of freedom.

	For explicit calculations, the exponential function containing the interaction operator must be expanded into a perturbation series. In the simplest case, this is just a Taylor expansion in the interaction strength which leads to the generating functional in $n$-th order of canonical perturbation theory

	\begin{align}\label{interacting_functional_pert_series}
		Z[ J, K]^{(n)} =& \left(1+ \ii\hat S_\mathrm I + \dots + \tfrac{\ii^n}{n!} \hat S_\mathrm I^n\right)  \int \d\Gamma_{\mathrm i}\, \nexp{\ii \int_{\mathrm i}^{\mathrm f} \d t \,  J_a\bar\varphi_a} \;.
	\end{align}
	Note that the expansion in orders of the interaction operator is the only approximation in the derivation of the generating functional from the classical equation of motion.

	From the generating functional, expectation values for different degrees of freedom can be extracted by functional derivatives with respect to $J_a$ and $K_a$.
	For example, a general field correlator reads
	\begin{align}
		&\langle {\varphi}_{j_1}(x_1)\dots {\varphi}_{j_m}(x_m) \chi_{j_{m+1}}(x_{m+1})\dots \chi_{j_{m+n}}(x_{m+1}) \rangle \\
		&= \der{}{J_{j_1}(x_1)}\cdots\der{}{J_{j_m}( x_m)}\cdots \der{}{K_{j_{m+n}}(x_{m+n})} Z[J,K]\, \bigg|_{J = 0 =K}\nonumber\;.
	\end{align}
	The derivatives extract the respective time-evolved field components from the generating functional which are then averaged over the initial field configuration, corresponding to an ensemble average.

	As the time evolution of degrees of freedom does not generally result in an equilibrium state, the above expression corresponds to a non-equilibrium ensemble average at times specified in the functional derivatives.

	While the generating functional is built upon microscopic degrees of freedom and their equations of motion, it is possible to extract information about macroscopic collective fields by means of operators. This is an important advantage of this theory.

	An example for such a collective field is the spatial number density $\rho$, if the degrees of freedom of the ensemble are specified to the phase space coordinates of $N$ particles (see for example \cite{2016NJPh...18d3020B} or \cite{2015PhRvE..91f2120V}). Then, a collective spatial particle density field at position $\vec q$ and time $t$ is simply the sum over all point-particle contributions at this position and time
	\begin{align}
	\rho(t, \vec q) = \sum_j \dirac(\vec q - \vec q_j(t))\;.
	\end{align}
	Fourier transforming this expression and replacing the particle positions by derivatives with respect to the conjugate source field, $\vec q_i(t) \rightarrow \der{}{J_{q_j}(t)}$, yields the particle density operator in Fourier space
	\begin{align} \label{denstiy_operator_fourier}
	& \hat \rho(t, \vec k) = \sum_j \hat \rho_j (t, \vec k) = \sum_j \nexp{-i\vec k \cdot \der{}{J_{q_j}(t)} }\;.
	\end{align}
	We will later need several other collective fields like velocity-density or pressure, which we will introduce as needed.

\section{Incorporating Fluid Dynamics} \label{incorporating_fluid_dynamics}

    The full theory of KFT in its exact form of Eq.~\eqref{general_interacting_functional} contains the complete dynamics that can arise in a particle ensemble, including the dynamics of particles in a fluid. However, performing actual calculations is complicated due to the nature of the theory: KFT describes the linear propagation of particles exactly, but in most applications interactions must be included in a perturbation series in the interaction operator. Ensembles dominated by particle propagation with only small disturbances due to interactions can be well represented, as they can be described in a low perturbative order.
    
    In contrast, a fluid is dominated by the interactions of microscopic particles: For ideal hydrodynamics, microscopic particles are assumed to be so tightly coupled that their mean free path is zero. Viscous effects arise when this assumption is relaxed to a finite but still small mean free path. To reach the regime of fluid dynamics in KFT, one would need to go to very high order in perturbation theory. Unfortunately, this is impossible for any practical calculation.
    
    Here, we nonetheless attempt to incorporate fluid dynamics using an idea close to the conventional approach to hydrodynamics: One of the assumptions in the conventional derivation is the existence of the hydrodynamical scale hierarchy. It states that one can choose a scale in such a way that it is much larger than the mean free path of particles and at the same time much smaller than the characteristic scale of the phenomena one is interested in. On this intermediate scale, fluid elements containing many microscopic particles can be introduced. Their properties are then defined as average quantities of all microscopic particles contained.
    
    This has two advantages: First, the frequent collisions of the microscopic particles quickly establish local thermodynamic equilibrium on the scale of each individual fluid element. This, in turn, allows defining local state variables like pressure and energy density.
    Also, microscopic conservation of particle number, energy, and momentum gives rise to macroscopic conservation laws such that further details of the microscopic interactions are rendered unimportant for the dynamics of the ensemble. 
    
    The time evolution of each fluid element is well described by a propagation with the average velocity and perturbations caused by accelerations due to pressure gradients and pressure-volume work. As long as no large gradients in the pressure and velocity field appear, these perturbations stay small.
    For viscous hydrodynamics, additional effects caused by diffusion of energy and momentum occur, which in most circumstances are only small perturbations as well.
    
    Within the framework of 
    KFT, these simplified dynamics motivate an approach to hydrodynamics based on mesoscopic particles similar to fluid elements. These mesoscopic particles propagate with a locally averaged velocity. The propagation of the mesoscopic particles is described by a Green's function contained in a free functional. Any necessary interactions can then be included with the help of interaction operators.
    As long as all interactions remain sufficiently weak, i.e.\ as long as the fluid does not display large gradients, it is likely that a low-order canonical perturbation theory already captures fluid dynamics.
    
    In the following sections, we set up a free generating functional and interaction operators that model both ideal and viscous hydrodynamics based on mesoscopic particles of equal mass. Each of these particles can be characterised by three properties: the first two are its centre of mass position $\vec q$ and velocity $\vec u$ (or equivalently momentum $\vec p$). As this velocity is defined as an average over the velocities of all microscopic particles contained by the mesoscopic particle, it will depend on the scale on which the mesoscopic particles are defined. The average velocity describes only part of the information that is originally contained in the velocities of the microscopic particles. Further information is contained in the velocity dispersion which we include here in the form of a local stress-energy tensor. In summary, we will describe fluid dynamics in KFT by introducing fluid particles, that propagate with the local average velocity field and carry with them a local representation of the stress-energy tensor.
    
    For the specific case of an isotropic fluid, the stress-energy tensor is fully characterised by the pressure $P$. If one assumes in addition, that the pressure is only caused by the random velocities of the particles and that the particles do not have any internal degrees of freedom, the pressure relates to the internal energy-density as $P = \frac{2}{3} \epsilon$.\\
    It is convenient to combine both pressure and internal energy-density into the enthalpy-density
    \begin{align}
    h = \epsilon + P = \tfrac{5}{3} \epsilon =  \tfrac{5}{2} P\;,
    \end{align}
    as this allows to include all necessary information about a fluid particle by assigning the enthalpy $\mathcal{H}$ as a third property to the mesoscopic particles.
    
	We emphasise again that in principle, it must be possible to derive these three properties as well as all interactions of hydrodynamics from a fully microscopic theory but this remains a task for the future. In an approach similar to smoothed particle hydrodynamics, we start from macroscopic hydrodynamical equations and construct an appropriate free theory and interaction operators to model hydrodynamics based on the mesoscopic particles described above. Thereby, we demand the existence of the hydrodynamical scale hierarchy to assume that the mesoscopic particles have no significant spatial extent in comparison with the fluid phenomena we are interested in and thus can effectively be modelled as point-like.

	\subsection{Free theory for mesoscopic particles} \label{free_theory_mesoscopic_particles}
	In the free theory of hydrodynamics, we include all dynamics that can be described as the propagation of mesoscopic particles. Namely, we set up an ensemble of equal-mass particles which are characterised by three properties: their centre of mass  $\vec q$, a centre of mass momentum $\vec p$, and an enthalpy $\mathcal H$.
				
	The particles change their position according to the propagation with their momentum. Both momentum and enthalpy of each individual particle are conserved and are carried around with the particle. Hence, the momentum-density or enthalpy-density of the fluid might change locally as particles converge or diverge.
				
	The dynamics of both ideal and viscous hydrodynamics, i.e.\ acceleration due to pressure gradients, pressure-volume work, and diffusive effects will later be included with the help of interaction operators. These effects will change the momentum and enthalpy of individual particles.
				

	To set up the free generating functional with the above dynamics, we first bundle the degrees of freedom of the $j$-th particle into a single vector
	\begin{align}
		\varphi_j(t) = \left(\vec q_j(t), \vec p_j(t), {\mathcal H}_j(t)\right)^\ntop\;,
	\end{align}
	with a corresponding source-field vector defined as
	\begin{align}
	J_j(t) = \left(\vec{J}_{q_j}(t), \vec{J}	_{p_j}(t), J_{{\mathcal H}_j}(t)\right)^\ntop \,.
	\end{align}
	For the free theory, we set all forces to zero. Hence the position of each particle only changes due to propagation, i.e.\ linearly with its momentum
	\begin{align}
	\dot{ \vec q}_j(t) = \frac{\vec p_j(t)}{m}\;,
	\end{align}
	and the particle's momentum and enthalpy are both conserved
	\begin{align}
	\quad \dot{\vec p}_j(t) = 0\; , \quad \dot{{\mathcal H}}_j(t) = 0 \;.
	\end{align}

	These equations of motion are solved by the retarded Green's function
	\begin{align}\label{hydro_greens_functions}
	G_R(t,t_{\mathrm i}) &=
	\left(
	\begin{array}{ccc}
	\mathcal I_3 &  \frac{t- t_{\mathrm i}}{m} \cdot \mathcal I_3 & 0\\
	0 &  \mathcal I_3 &  0\\
	0 &  0& 1
	\end{array}
	\right)\theta(t-t_{\mathrm i})\\
	&\eqqcolon \left(\begin{array}{lll}
	g_{qq}(t,t_{\mathrm i}) & g_{qp}(t,t_{\mathrm i})  & g_{q{\mathcal H}}(t,t_{\mathrm i})\\
	g_{pq}(t,t_{\mathrm i})  & g_{pp}(t,t_{\mathrm i}) & g_{p{\mathcal H}} (t,t_{\mathrm i})\\
	g_{{\mathcal H}q}(t,t_{\mathrm i}) & g_{{\mathcal H}p}(t,t_{\mathrm i}) & g_{{\mathcal H}{\mathcal H}}(t,t_{\mathrm i})
	\end{array}
	\right)\theta(t-t_{\mathrm i})\;,
	\end{align}
	where $\mathcal I_3$ is the three dimensional unit matrix.

	Using the above Green's functions as well as Eq.~\eqref{general_interacting_functional}, we can define the free generating functional for mesoscopic particles as

	\begin{align}\label{mesoscopic_generating_free_functional}
	\tilde Z_0\left[  \tens J, \tens K\right] = \int \d\Gamma_{\mathrm i}\, \nexp{\ii \int_{\mathrm i}^{\mathrm f} \d t \, \langle\tens J (t), \bar{\boldsymbol \varphi}(t)\rangle }\;,
	\end{align}
	where the tilde marks the free generating functional specified for mesoscopic particles. 
	For the sake of compact notation we bundle the respective quantities of all $N$ particles into a single tensorial object which is denoted by a boldface character. Defining the $N$-dimensional column vector $\vec e_j$, whose only non-vanishing entry is unity at component $j$, these read
	\begin{align}
	\tens J &:= \vec  J_j \otimes \vec e_j\;, \qquad\bar {\boldsymbol \varphi}:= \bar {\varphi}_j \otimes \vec e_j\;, \qquad \tens K := \vec K_j \otimes \vec e_j\;,
	\end{align}
	with a scalar product defined by
	\begin{equation}
	\langle\tens a,\tens b\rangle := \left(\vec a_j\otimes\vec e_j\right)\left(\vec b_k\otimes\vec e_k\right) =
	\vec a_j\cdot\vec b_k\,\delta_{jk} = \vec a_j\cdot\vec b_j\;,
	\end{equation}
	where a sum over $j$ is implied.
	In any other context, the angular brackets indicate the non-equilibrium ensemble average described in the last section.

	The time-evolved degrees of freedom of the $j$-th particle $\bar \varphi_j(t)$  read
	\begin{equation}\label{hydro_bar_phi}
	\bar \varphi_j(t) =G_\mathrm R (t,t_\mathrm i)\, \varphi_j^{\,\mathrm{(i)}} - \int_{t_\mathrm{i}}^t\d t'\,G_\mathrm R(t,t')\vec K_j(t') \;,
	\end{equation}
	with the additional source field
	\begin{align}
	\vec K_j(t) = \left( \vec K_{q_j}(t), \vec K_{p_j}(t),K_{{\mathcal H}_j}(t) \right)^\ntop \;.
	\end{align}
	Finally, the $\d \Gamma_{\mathrm i}$-integration runs over all initial degrees of freedom weighted by their initial probability distribution
	\begin{align}
	\d \Gamma_\mathrm i = P\left(\tens q^{(\mathrm i)}, \tens p^{(\mathrm i)},  \mathcal H  \hspace{-3.5mm} \mathcal H ^{(\mathrm i)}\right)\; \d^{3N} q^{(\mathrm i)} \; \d^{3N} p^{(\mathrm i)} \; \d^{N} \mathcal H^{(\mathrm i)}\;,
	\end{align}
	where the probability distribution is chosen such that it samples the smooth macroscopic fields of a fluid.

	Note that this choice of dynamics specifies the ensemble to a fluid with isotropic pressure. For the non-isotropic case, the full stress-energy tensor would be needed instead of $\mathcal H$. However, this would render the notation too involved for the purpose of this paper.

	\subsection{Collective fields} \label{collective_fields}
	To include appropriate interactions between mesoscopic particles, we first need to define collective field operators for number-density, velocity-density, momentum-density, and energy-density as well as pressure.

	We can define the operator for the number-density contribution of the $j$-th particle in configuration space as
	\begin{align}
	\hat \rho_j(t,\vec q)& =  \dirac\left(\vec q - \der{}{J_{q_j}(t)}\right) \\
		&= \int \frac{\d^3 k}{(2 \pi)^3} \, \exp\left\{- \ii \vec k \cdot\left( \der{}{J_{q_j}(t)}-\vec q\right) \right\}\;, \nonumber
	\end{align}
	the operators for velocity and momentum of the $j$-th particle by 
	\begin{align*}
	\hat{\vec u}_j(t) = \frac{\vec p_j}{m } = \frac{1}{m} \der{}{J_{p_j}(t)}\;, \qquad
	\o_{\vec p_j}(t)  = \vec p_j =  \der{}{J_{p_j}(t)}\;,
	\end{align*}
	and the operator for the enthalpy of the $j$-th particle by
	\begin{align}
	&\hat{\mathcal H}_j (t) = \der{}{J_{\mathcal H_j}(t)}\;.
	\end{align}
	As a consequence of the particle-based description, the velocity operator, as well as the momentum and enthalpy operators can never describe the spatial dependence of a collective field in configuration space. This is reflected by the absence of any spatial coordinate in their definition.
	Therefore, to obtain macroscopic collective information about these quantities in configuration space, they must be paired with a density operator carrying the same particle index (for more details see  \cite{2015PhRvE..91f2120V}).
	While we introduce the above `naked' operators for the sake of an easier notation, they must always appear as a velocity-density, momentum-density or enthalpy-density,
	\begin{align}
	\hat{ \rho \vec u}_j(t, \vec q) &= \hat \rho_j(t,\vec q) \hat{\vec u}_j(t)\;,\\
	\hat{\rho \vec p}_j(t, \vec q) &= \hat\rho_j(t,\vec q) \hat{\vec p}_j(t) \;,\\
	\hat{\rho \mathcal H}_j(t, \vec q) &= \hat\rho_j(t,\vec q) \hat{\mathcal H}_j(t)\;.
	\end{align}
	Using the enthalpy-density, we can finally define operators for pressure and internal energy
	\begin{align}
	& \hat{\varepsilon}_j(t,\vec q\,) = \tfrac{3}{5} \hat{\mathcal H}_j(t)\, \hat\rho_j(t,\vec q)\;, \\
	&\hat P_j(t,\vec q\,) = \tfrac{2}{5} \hat{\mathcal H}_j(t) \, \hat{\rho}_j(t,\vec q) \;.
	\end{align}
	Note that the above operators extract properties of a single particle (the $j$-th particle). The corresponding macroscopic fields are found by summing over all particles.

	\subsection{Interactions} \label{subsection_SPH_interactions}

	The free theory set up in the previous section already contains the propagation of mesoscopic particles and the corresponding transport of particle properties.

	To arrive at a fluid-like behaviour of the ensemble, we need to incorporate all remaining dynamics with the help of interaction operators. These need to contain accelerations by pressure gradients and pressure-volume work in the case of ideal hydrodynamics and additional effects of momentum and energy diffusion for viscous hydrodynamics.

	The main difficulty in the derivation of these interaction operators is the fundamental difference between conventional hydrodynamics and KFT: The hydrodynamical interactions we are looking for depend on derivatives of continuous fields; KFT, however, is a particle-based approach. By adding up particle contributions, it is possible to form fields. However, as particles are point-like, these fields are discontinuous and their local derivatives are ill-defined.

	Here, we present two possible derivations that deal with this difficulty and arrive at proper interaction operators. The first is closely linked to smoothed particle hydrodynamics (SPH~\cite{SPH-review}): With the help of a density kernel, discrete particles and their properties are smoothed into continuous fields. The interactions for these fields are read off the hydrodynamical equations and are rediscretised to describe the dynamics of individual particles.

	The second derivation follows heuristic arguments that interpret the physical processes taking place on the scale of the mesoscopic particles.

	Both approaches have their merits and lead to interaction terms that only differ in irrelevant details. As the approaches highlight different aspects of the properties and interpretations of the hydrodynamical model, we believe it to be advantageous to present both. The mathematically more rigorous and elegant SPH-derivation is described in the subsequent paragraphs, while the heuristic derivation can be found in Appendix~\ref{appendix_heuristic_derivation}.

	SPH translates the smooth field equations of conventional hydrodynamics to a particle picture. This is done by smoothing out each particle by a three-dimensional normalised kernel $v(\vec q - \vec q_i)$ centred on the position of the original point-particle $\vec q_i$.

	We adopt this approach and rewrite density, momentum-density, velocity-density, pressure, energy-density, and pressure flux in terms of smoothed particles
	\begin{align}
		&&\label{density_disc}
		\rho(\vec q) &\quad \rightarrow \quad  \sum_i v(\vec q-\vec q_i)\,, \\
		&& \label{momentum_disc}
		\rho \vec p (\vec q) & \quad \rightarrow \quad \sum_i \vec p_i \, v(\vec q-\vec q_i)\,, \\
		&& \label{velocity_disc}
		\rho \vec u (\vec q) & \quad \rightarrow \quad \sum_i \vec u_i \, v(\vec q-\vec q_i) \, ,\\
		&& \label{pressure_disc}
		P(\vec q) = \tfrac{2}{5} \mathcal{H} \rho (\vec q) &\quad \rightarrow \quad\sum_i \tfrac{2}{5} \mathcal{H}_i\, v(\vec q- \vec q_i) \,,\\
		&& \label{energy_disc}\epsilon(\vec q) = \tfrac{3}{5} \mathcal{H} \rho (\vec q)  & \quad \rightarrow \quad \sum_i \tfrac{3}{5} \mathcal{H}_i\, v(\vec q- \vec q_i)\,, \\
		&& \label{pressure_vel_disc}P\vec u(\vec q) = \tfrac{2}{5} \mathcal{H} \vec u  \rho (\vec q) & \quad \rightarrow \quad \sum_i \tfrac{2}{5} \mathcal{H}_i \, \vec u_i \, v(\vec q- \vec q_i)\,,
	\end{align}
	where the sums run over all particles. With these fields at hand we can construct the KFT interaction operators.

	\subsubsection{Acceleration by pressure gradients}\label{SPH_acceleratrion_by_pressure_gradients}
	To model acceleration due to pressure gradients, we start with the time evolution of the momentum field as it appears in the Euler equation
	\begin{align}
		\dot p (\vec q_1)& = - \frac{\partial_{q_1} P(\vec q_1)}{\rho(\vec q_1)} = - \frac{1}{\rho(\vec q_1)} \sum_i \tfrac{2}{5} \mathcal H_i  \,   \partial_{q_1} \, v(\vec q_1-\vec q_i)\;,
	\end{align}
	where we inserted the discretised pressure field Eq.~\eqref{pressure_disc}.

	We then pull the position of the $i$-th particle into a Dirac delta distribution that we identify with the density of the $i$-th particle,
	\begin{align}
		\dot p (\vec q_1)&=  - \frac{1}{\rho(\vec q_1)}   \sum_i \int \mathrm d^3  q_2 \, \tfrac{2}{5}\mathcal H_i  \, \dirac(\vec q_2-\vec q_i) \, \partial_{q_1}\, v(\vec q_1-\vec q_2) \nonumber \\
		&=  - \frac{1}{\rho(\vec q_1)}  \sum_i \int \mathrm d^3 q_2 \,\tfrac{2}{5} \mathcal H_i \, \rho_i(\vec q_2) \,  \partial_{q_1} \, v(\vec q_1-\vec q_2)\;.
	\end{align}

	The momentum change at position $\vec q_1$ must be reassigned to particles at this position according to their (spatial) contribution. To this end, we weight the momentum change at $\vec q_1$ with the density contribution of the $j$-th particle and integrate over the entire space,
	\begin{align} \label{SPH_momentum_change_ideal_hydro}
		\dot p_j &= \int \mathrm d^3 \vec q_1 \, \dot p(\vec q_1)\, \rho_j(\vec q_1) \\
		&=   -  \tfrac{2}{5}  \sum_i \int \mathrm d^3 q_2\,\mathrm d^3 q_1 \, \frac{1}{\rho(\vec q_1)}  \,\mathcal H_i \, \rho_i(\vec q_2) \, \rho_j(\vec q_1) \,  \partial_{q_1}  \, v(\vec q_1-\vec q_2) \nonumber\;.
	\end{align}


	\subsubsection{Pressure volume work}
	To include pressure-volume work in an analogous way, we first determine the time evolution of the enthalpy from the energy conservation equation of ideal hydrodynamics,
	\begin{align}
	\dot{\mathcal H } (\vec q_1)= \frac{5}{3}\frac{\dot \epsilon(\vec q_1)}{\rho(\vec q_1)} = - \frac{5}{3}\frac{P(\vec q_1)}{\rho(\vec q_1)}\, \partial_{q_1} \, \vec u(\vec q_1)\;.
	\end{align}
	The field $\vec u$ cannot be discretised on its own. To circumvent this problem, we can rewrite the derivative as
	\begin{align} \label{velocity_discretisation_1}
	\dot{\mathcal H } (\vec q_1) &= - \frac{5}{3} \, \frac{1}{\rho(\vec q_1)}\, \left[\partial_{q_1}  \,P\vec u(\vec q_1)- u(\vec q_1)\, \partial_{q_1}  P(\vec q_1)\right]\;.
	\end{align}
	
	Inserting Eq.~\eqref{pressure_disc} and Eq.~\eqref{pressure_vel_disc} into Eq.~\eqref{velocity_discretisation_1}, we get
	\begin{align}
	&\dot{\mathcal H } (\vec q_1) = - \frac{2}{3} \, \frac{1}{\rho(\vec q_1)}\, \sum_i \, \mathcal H_i\, \left[ \vec u_i - \vec u(\vec q_1)\right] \partial_{q_1} v(\vec q_1-\vec q_i)\\
	&\vspace{4mm}= - \frac{2}{3} \, \frac{1}{\rho(\vec q_1)}\, \sum_i \int \mathrm d^3 q_2 \, \mathcal H_i\,  \left[ \vec u_i - \vec u(\vec q_1)\right] \rho_i(\vec q_2) \, \partial_{q_1} v(\vec q_1-\vec q_2) \nonumber\;,
	\end{align}
	where in the second line we again introduced a Dirac delta distribution $\dirac(\vec q_2-\vec q_i)$ which we identify with the density contribution of the $i$-th particle at position $\vec q_2$.
	
	The enthalpy change is then represented by particles. Picking out the enthalpy change of the $j$-th particle, we get
	
	\begin{align} \label{SPH_enthalpy_change_ideal_hydro}
	\dot{\mathcal H }_j &= \int \mathrm d^3 q_1 \, \rho_j(\vec q_1) \, \dot{\mathcal H}(\vec q_1) \nonumber\\
	&= - \frac{2}{3} \, \sum_i \int \mathrm d^3  q_2 \, \mathrm d^3  q_1 \,  \frac{\mathcal H_i}{\rho(\vec q_1)}\, \left[  \vec u_i - \vec u_j\right]  \\
	&\hspace{2.8cm}  \times \, \rho_j(\vec q_1) \, \rho_i(\vec q_2) \, \partial_{q_1} v(\vec q_1-\vec q_2) \nonumber\;.
	\end{align}
	Here, the velocity field at position $\vec q_1$ was replaced with the velocity of the $j$th particle. This can be done since velocities are sampled from a smooth field and the appearing $\rho_j(\vec q_1)$ specifies the position of the $j$th particle to be $\vec q_1$. 
	

	\subsubsection{Diffusive effects}
	To find an interaction operator for diffusive effects, we first define diffusive currents of momentum and energy-density and then model their impact on the properties of the mesoscopic particles. Diffusive currents are proportional to gradients in the corresponding densities. For the energy-density and momentum-density currents we get
	\begin{align} \label{energy_current} 
	\vec j (\vec q_1) & = -\mu \, \partial_{q_1} \epsilon(\vec q_1) \;,  \\
	\label{stress_energy_tensor} 
	T^{ab} (\vec q_1) & = - c_{ab} \left(\partial_{q_1}^a\, \rho p^b (\vec q_1) + \partial_{q_1}^b\, \rho p^a (\vec q_1) \right) \nonumber \\
	&\eqqcolon - c_{ab} \left( \partial_{q_1}^a\, \rho\, p^b (\vec q_1)\right)_\text{s}\;,
	\end{align}
	where the indices $a$ and $b$ denote vector components such that $\partial_{q_1}^a$ implies the derivative with respect to the $a$-th component of the position vector $\vec q_1$ and $ \rho p^b (\vec q_1)$ is the $b$-th component of the momentum density field at $\vec q_1$. 
	$\mu$ and $c_{ab}$ are diffusive constants that depend, among other things, on the microscopic particles' mean free path and collision rates and have the unit $[\mu] = [c_{ab}] = \mathrm{\tfrac{m^2}{s}}$. The minus signs ensure that diffusive currents flow from over-dense to under-dense regions and $(\dots)_\text{s}$ denotes the symmetrization of the  stress-energy tensor $T^{ab}$. This is needed for momentum and angular momentum conservation. More precisely, it ensures that no viscous effects arise for solid body rotation. For a proof see for example~\cite{BartelmannBuch}.
	
	In the case of isotropic diffusion, only two independent momentum diffusion constants remain
	\begin{align} \label{diffusion_constants}
	c_{ab} =  \left\{
	\begin{array}{ll}
	\zeta,& a = b \\
	\nu, &  a \neq b
	\end{array}\right. \;,
	\end{align}
	where $\zeta$ describes the diffusion in the direction of the average flow (bulk flow), and $\nu$ the diffusion orthogonal to it (shear flow).
	
	
	Divergences in the diffusive currents will lead to a local change of enthalpy and momentum. Using the discretization Eq.~\eqref{momentum_disc}, we get for the momentum change,
	\begin{align}
	\dot{p}^a (\vec q_1) &= - \frac{1}{\rho(\vec q_1)}\, \partial_{ q_1}^b \, T^{ab} (\vec q_1)  \\
	&=\frac{c_{ab}}{\rho(\vec q_1)}\,\sum_i  \left( p_i^b\, \partial_{q_1}^a \right)_\text{s}  \partial_{ q_1}^b v(\vec q_1- \vec q_i) \nonumber\\
	& = \frac{c_{ab}}{\rho(\vec q_1)}\,\sum_i \int \mathrm d^3 q_2 \,\rho_i(\vec q_2)  \, \left( p_i^b\, \partial_{q_1}^a \right)_\text{s}  \partial_{ q_1}^b v(\vec q_1- \vec q_2) \nonumber\;,
	\end{align}
	where the indices $a$ and $b$ indicate vector components and a summation over $b$ is implied. In the last line a Dirac delta distribution was used to pull $\vec q_i$ from the kernel and the delta distribution was identified with the contribution of the $i$-th particle.
	
	The momentum change of the field is converted to the momentum change of the $j$-th particle by
	\begin{align} \label{SPH_momentum_change_viscous}
	\dot{p}^a_j &= \int \mathrm d^3 q_1 \, \rho_j(\vec q_1)\,  \dot{p}^a (\vec q_1) \nonumber \\
	&=  c_{ab}\,\sum_i \int  \mathrm d^3 q_1 \,\mathrm d^3 q_2 \, \frac{\rho_j(\vec q_1)\, \rho_i(\vec q_2)}{\rho(\vec q_1)}  \\
	&\qquad \times \left( p_i^b\, \partial_{q_1}^a \right)_\text{s}  \partial_{ q_1}^b v(\vec q_1-\vec q_2) \nonumber\;.
	\end{align}

	For the enthalpy change, we need to consider an additional phenomenon: the momentum-density current tensor has the unit of pressure and hence performs pressure-volume work. Including this effect in addition to the changes caused by divergences in the energy-density current, the enthalpy change is
	\begin{align} \label{enthalpy_change_viscous_non-discretized}
	\dot{\mathcal H} (\vec q_1) &= - \tfrac{5}{3}\frac{1}{\rho(\vec q_1)} \partial_{q_1} \cdot \vec j (\vec q_1) -  \tfrac{5}{3} \frac{1}{\rho(\vec q_1)}\, T^{a b}(\vec q_1)\,  \partial_{q_1}^a u^b(\vec q_1)\;.
	\end{align}
	Once again, the velocity field cannot be discretized on its own. We rewrite
	\begin{align}
	T^{a b}(\vec q_1)\,\partial_{q_1}^a u^b(\vec q_1) 
	&=   \partial_{q_1}^a \, T^{a b}u^b(\vec q_1)\,  -  u^b(\vec q_1) \, \partial_{q_1}^a \, T^{a b}(\vec q_1) \;,
	\end{align}
	with the discretization 
	\begin{align} \label{discretized_viscous_pressure_volume_work}
		T^{a b}(\vec q_1)\,\partial_{q_1}^a u^b(\vec q_1) 
		=   - & c_{ab} \sum_i  \left[u_i^b   -  u^b(\vec q_1) \right] \, \\
		  & \times \left( p^b_i \, \partial_{q_1}^a\right)_\text{s} \,\partial_{q_1}^a \,v(\vec q_1-\vec q_i) 
		\;, \nonumber
	\end{align}

Inserting Eq.~\eqref{discretized_viscous_pressure_volume_work} as well as Eq.~\eqref{energy_current} with the discretization Eq.~\eqref{energy_disc} into Eq.~\eqref{enthalpy_change_viscous_non-discretized} yields
	\begin{align}
	\dot{\mathcal H}(\vec q_1) &=  \frac{\mu}{\rho(\vec q_1)} \,  \sum_i  \mathcal H_i  \, \partial_{q_1}^2  v(\vec{q}_1-\vec{q}_i)\\
	 & +  \tfrac{5}{3}  \frac{c_{ab}}{\rho(\vec q_1)} \sum_i  \left[u_i^b   -  u^b(\vec q_1) \right] \, \left( p^b_i \, \partial_{q_1}^a\right)_\text{s} \,\partial_{q_1}^a \,v(\vec q_1-\vec q_i)  \nonumber\;.
	\end{align}
	Introducing Dirac delta distributions and identifying them with the one-particle density contributions yields
	\begin{align} 
	\dot{\mathcal H}(\vec q_1) &=  \frac{\mu}{\rho(\vec q_1)} \,  \sum_i \int \mathrm d^3 q_2  \, \rho_i(\vec q_2) \,  \mathcal H_i  \, \partial_{q_1}^2  v(\vec{q}_1-\vec{q}_2)\\
	& +  \tfrac{5}{3}  \frac{c_{ab}}{\rho(\vec q_1)} \sum_i   \int \mathrm d^3 q_2 \, \rho_i(\vec q_2) \,\left[u_i^b   -  u^b(\vec q_1) \right]\nonumber \\
	& \phantom{+} \times \left( p^b_i \, \partial_{q_1}^a\right)_\text{s} \,\partial_{q_1}^a  \,v(\vec q_1-\vec q_2)  \nonumber\;.
	\end{align}
	In a last step, we convert the enthalpy change of the field to the enthalpy change of the $j$-th particle and use that velocities are sampled from a smooth velocity field to replace the velocity at position $\vec q_1$ with the velocity of the $j$-th particle,
	\begin{align} \label{SPH_enthalpy_change_viscous}
	\dot{\mathcal H}_j &= \mu \,  \sum_i \int \mathrm d^3 q_1\, \mathrm d^3 q_2 \,  \frac{\rho_j(\vec q_1)\, \rho_i(\vec q_2) }{\rho(\vec q_1)} \, \mathcal H_i  \, \partial_{q_1}^2  v(\vec q_1- \vec q_2) \nonumber\\
		& +  \tfrac{5}{3} c_{ab}  \sum_i  \int \mathrm d^3 q_1\, \mathrm d^3 q_2 \,  \frac{\rho_j(\vec q_1)\, \rho_i(\vec q_2) }{\rho(\vec q_1)}\,\left[u_i^b   -  u^b_j \right] \,\nonumber \\
		& \phantom{+}\times  \left( p^b_i \, \partial_{q_1}^a\right)_\text{s} \,\partial_{q_1}^a  \,v(\vec q_1-\vec q_2) \;.
	\end{align}

	\subsection{Interaction operators for ideal and viscous hydrodynamics} \label{Interaction_operators_for_ideal-and_viscous_hydrodynamics}
	As described in the introduction to KFT in section \ref{summary_KFT}, the interaction term is defined as
	 \begin{align}
	 S_I&=\int \d t \left[{ E}_{\mathrm I,p}(t) \cdot  \chi_{p_j} (t)+E_{\mathrm I , \mathcal H} (t)\,\chi_{\mathcal H_j} (t)\right] \nonumber \\
	 & = - \sum_j \int \d t \left[{\dot p_j}(t) \cdot  \chi_{p_j} (t)+\dot{ 	\mathcal H}_j (t)\,\chi_{\mathcal H_j} (t)\right]\;,
	 \end{align}
	 where $\dot{\mathcal H}_j (t)$ and $\dot p_j(t)$ in ideal hydrodynamics are given by Eqs.~\eqref{SPH_momentum_change_ideal_hydro} and Eq.~\eqref{SPH_enthalpy_change_ideal_hydro}, respectively.

	 To transform this object into an operator, all particle densities, velocities and enthalpies must be replaced by their respective operators. However, in both $\dot p_j$ and $\dot{\mathcal{H}}_j$ appears an inverse density. This object cannot be expressed by functional derivatives. 

	 We handle this complication by approximating the inverse densities in a Taylor series around the mean density $\bar \rho$ of the ensemble. In this paper we already truncate the approximation at $0$\textsuperscript{th} order, as this is fully sufficient for a proof of concept,
	 \begin{align}\label{approx_densitiy}
	 	\frac{1}{\rho(\vec q_1)} \approx \frac{1}{\bar \rho}\,.
	 \end{align}
	 For further comments on this approximation, we refer to the end of Section~\ref{Evolution_equations_of_macroscopic_fields}.

	Using this approximation, the interaction for ideal hydrodynamics takes the form
	\begin{align}
		S_I^\text{\tiny id} & =    \frac{2}{5 \bar \rho}  \sum_{(i,j)} \int \mathrm d t_1\, \mathrm d^3 q_1\, \mathrm  d^3 q_2\,   \rho_j(\vec q_1)\,   \rho_i(\vec q_2) \\
		& \hphantom{ \frac{2}{5 \bar \rho}  \sum_{(i,j)} \int}    \times\,\mathcal H_i \,  \left[ \partial_{q_1}  v(\vec q_1 - \vec q_2) \right] \,  \chi_{p_j} (t_1)  \,\nonumber\\
		&+ \frac{2}{3 \bar \rho} \, \sum_{(i,j)} \int \mathrm d t_1\, \mathrm d^3 q_1\, \mathrm  d^3 q_2\, \rho_j(\vec q_1)  \, \rho_i(\vec q_2) \nonumber \\
		&\hphantom{ \frac{2}{5 \bar \rho}  \sum_{(i,j)} \int}    \times \,\mathcal H_i\, \left[  \vec u_i - \vec u_j\right]  \, \left[ \partial_{q_1}  v(\vec q_1-\vec q_2) \right]\,  \chi_{\mathcal H_j} (t_1)\nonumber\;.
	\end{align}
	The notation $\sum_{(i,j)}$ denotes the sum over $i$ and $j$ with $i\neq j$ to exclude self-interactions. 
	
	The kernel ${v(\vec q_1 - \vec q_2)}/{\bar \rho}$ can be interpreted as the range of the interaction, which is perhaps most evident in the heuristic derivation in Appendix~\ref{appendix_heuristic_derivation}.

	Since conventional hydrodynamics is a local theory we expect to recover it in the limit of vanishing interaction range
	\begin{align}
	v(\vec q_1 - \vec q_2) \rightarrow \dirac(\vec q_1 - \vec q_2)\;.
	\end{align}
	In order to take this limit, we need to remove the derivative from the kernel by replacing
	\begin{align}
	\partial_{q_1}  v(\vec q_1-\vec q_2) = - \partial_{q_2}  v(\vec q_1-\vec q_2)
	\end{align}
	 and integrating by parts. The interaction then turns into
		\begin{align} \label{SPH_interaction_operator_ideal}
		S_I^\text{\tiny id} & =  \frac{2}{5 \bar \rho}  \sum_{(i,j)} \int \mathrm d1\, \rho_j(1)\,  \mathcal H_i  \,  \left[ \partial_{1}  \, \rho_i(1)\right] \, \chi_{p_j} (1)\,  \\
		&+  \frac{2}{3 \bar \rho} \, \sum_{(i,j)} \int \mathrm d1 \,\rho_j(1) \, \left[  \vec u_i - \vec u_j\right] \,\mathcal H_i\, \left[\partial_{1}  \, \rho_i(1) \right]\,  \chi_{\mathcal H_j} (1) \nonumber\;,
		\end{align}
	where we used the abbreviations
	\begin{align} \label{realspace_abbreviations}
	&\mathrm d1 = \mathrm d^3 q_1 \mathrm d t_1 \qquad  \partial_{1} = \partial_{q_1} \\
	&\rho(1) = \rho(t_1,\vec q_1) \qquad \rho(2) = \rho(t_2, \vec q_2)\nonumber \\
	&\chi_{p_j} (1) = \chi_{p_j}(t_1) \qquad \chi_{\mathcal H_j} (1) = \chi_{\mathcal H_j}(t_1)\;. \nonumber
	\end{align}
	Finally, the expression for the interaction can be turned into an operator by replacing $\chi_{p_j} (1) $ and $\chi_{\mathcal H_j} (1) $ by their respective functional $K$-derivatives and all particle densities, momenta, and enthalpies by the corresponding operators described in Section~\ref{collective_fields}.

	To add diffusive effects, we take the diffusive momentum and energy transport from Eq.~\eqref{SPH_momentum_change_viscous}, and Eq.~\eqref{SPH_enthalpy_change_viscous} and proceed analogously:  Taking the approximation of the inverse densities Eq.~\eqref{approx_densitiy}, further taking the limit of vanishing interaction ranges and moving the derivative from the kernel to the density contribution of the $i$-th particle, we arrive at
		\begin{align}\label{SPH_interaction_operator_viscous}
		& S_I^\text{\tiny dif} = - \, \frac{c_{ab}}{\bar \rho}\,\sum_{(i,j)} \int  \mathrm d1 \, \rho_j(1)\,  \left[\left( p_i^b\, \partial_{1}^a \right)_\text{s}  \partial_{1}^b \, \rho_i(1)\right]\,\chi_{p_j}^a (1) \\
		&- \frac{\mu}{\bar \rho}  \,  \sum_{(i,j)} \int \mathrm d1
		\, \rho_j(1) \,  \mathcal H_i   \, \left[ \partial_{1} ^2\, \rho_i(1)\right]\,\chi_{\mathcal H_j} (1) \,  \nonumber\\
		& - \tfrac{5}{3} \frac{c_{ab}}{\bar \rho} \sum_{(i,j)} \int \hspace{-1mm} \mathrm d 1\, \rho_j(1) \left[u_i^b   -  u^b_j \right]\left[ \left( p^b_i \, \partial_{1}^a\right)_\text{s} \partial_{1}^a \, \rho_i(1)\right] \,\chi_{\mathcal H_j} (1) . \nonumber
		\end{align}
		with the abbreviations from Eq.~\eqref{realspace_abbreviations}.
	To arrive at the interaction operators, all appearing positions, momenta and conjugate source fields must again be replaced by their respective operator expressions as described for ideal hydrodynamics.

	Adding the interaction operators for ideal hydrodynamics and diffusive effects finally yields the full interaction operator for viscous hydrodynamics
	\begin{align}
	\hat S_I^\text{\tiny vis} = \hat S_I^\text{\tiny id} + \hat S_I^\text{\tiny dif} \;.
	\end{align}

	The interaction operators for ideal and viscous hydrodynamics, together with the free theory described in section \ref{free_theory_mesoscopic_particles}, define the dynamics of a fluid within KFT. In the remainder of this paper, we will inspect the properties of this model. To this end, we derive evolution equations for macroscopic fields from mesoscopic dynamics in the following section \ref{Evolution_equations_of_macroscopic_fields} and present the results of an exemplary calculation in section \ref{example-calculation}.

\section{Evolution equations of macroscopic fields}
	\label{Evolution_equations_of_macroscopic_fields}

	Conventional hydrodynamics is expressed by evolution equations for macroscopic fields, namely the density, velocity and energy-density. In order to compare the behaviour of the KFT fluid model to conventional hydrodynamics, we derive the corresponding macroscopic evolution equations from the dynamics of the mesoscopic particles.

	In order to substantially abbreviate future calculations, it is beneficial to derive evolution equations for a more general interaction. The interactions for both ideal and viscous hydrodynamics can be brought into the general form
	\begin{align}\label{general_hydro_interaction}
	\hat S_\mathrm I =
	& \int \d t  \sum_j \vec E_{\mathrm I_{pj}}(\hat{\vec q}_j, \hat{\vec p}_j, \hat{{\mathcal H}}_j)  \cdot \hat{\vec \chi}_{p_j} \\
	+ &\int \d t  \sum_j  E_{\mathrm I_{\mathcal Hj}}(\hat{\vec q}_j, \hat{\vec p}_j, \hat{\mathcal H}_j) \, \hat{\chi}_{{\mathcal H}_j}\;,
	\end{align}
	where the hats indicate that all particle positions, momenta and enthalpies as well as the auxiliary fields $\vec \chi_p$ and $\chi_\mathcal H$ are replaced by their respective functional derivatives. The full interacting theory then is
	\begin{align}\label{interacting_functional_general_fluid}
	\tilde Z[H,\tens J,\tens K] = \e&^{\ii \hat S_{\mathrm I}} \, \tilde Z_0[H,\tens J ,\tens K] \\
	= \exp&\left\{\ii \int \d t  \sum_j {\vec E}_{\mathrm I_{pj}}(\hat{\vec q}_j, \hat{\vec p}_j, \vec{\mathcal H}_j) \cdot \hat{\vec \chi}_{p_j}\right.\nonumber\\
	\nonumber &\left.+ \ii \int \d t  \sum_j E_{\mathrm I_{\mathcal H j}}(\hat{\vec q}_j, \hat{\vec p}_j, \hat{\mathcal H}_j) \, \hat \chi_{{\mathcal H}_j}\right\} \,
	\tilde Z_0[H,\tens J ,\tens K]\nonumber\;.
	\end{align}
	For all practical calculations the exponential must be expanded into a Taylor series and truncated at a certain order
	\begin{align} \label{taylor_expansion_interaction_hydro}
		\tilde Z[H,\tens J,\tens K]^{(n)} = \left(1 + \ii \hat S_{\mathrm I} + \dots + \frac{\ii^n}{n!} \hat S_{\mathrm I}^n \right)\, \tilde Z_0[H,\tens J ,\tens K]\;.
	\end{align}
	Deriving macroscopic evolution equations from this general interaction is a long and formal calculation without much physical insight and can be found in Appendix~\ref{appendix_macroscopic_evolution}. Here, we just state the relevant evolution equations. These are
	\begin{align}
	& \partial_t \ave{\rho(t, \vec q )}^{(n)}  + \, \partial_{\vec q} \cdot\ave{\rho\vec u (t,\vec q)}^{(n)}\label{general_int_continuity} = 0\\
	& \partial_t \ave{\rho \vec u (t, \vec q)}^{(n)}  +\, \partial_{\vec q}\cdot \ave{\rho\vec u \otimes \vec u(t,\vec q)}^{(n)}  \label{general_int_euler}
	= - \ave{  \tfrac{1}{m}\vec E_{\mathrm I_p}  \rho(t, \vec q )} ^{(n-1)} \\
	& \partial_t \ave{\varepsilon(t, \vec q)}^{(n)}  + \,\partial_{\vec q} \cdot\ave{\varepsilon\vec u (t, \vec q) }^{(n)}\label{general_int_energy}
	=  - \ave{ \tfrac{3}{5} E_{\mathrm I_{\mathcal H}}\rho(t,\vec q )}^{(n-1) }\;.
	\end{align}
	where the superscripts in parentheses indicate the order in the taylor expansion Eq.~\eqref{taylor_expansion_interaction_hydro}. For example
	\begin{align}
	 \ave{ \tfrac{3}{5} E_{\mathrm I_{\mathcal H}}\rho(t,\vec q )}^{(n-1) }
		= & \tfrac{3}{5} \sum_j E_{\mathrm I_{\mathcal Hj}}(\hat{\vec q}_j, \hat{\vec p}_j, \vec{\mathcal H}_j) \, \hat \rho_j(t,\vec q ) \\
		&\times \frac{\ii^{n-1}}{(n-1)!} \hat S_{\mathrm I}^{n-1} \,\tilde Z_0[H,\tens J ,\tens K]\biggr\rvert_{0} \;.\nonumber
	\end{align}

	 Note that the convection terms are of the same perturbation order as the time derivatives on the left-hand side; the particles' advection with the fluid stream is described exactly. All other terms are reduced by one perturbation order. Hence, all effects included via interaction operators can only be described exactly in arbitrarily high perturbation order; the series expansion is only able to approximate the full dynamics.

	\subsection{Terms of ideal hydrodynamics} \label{macroscopic_evol_ideal_hydro}
	The evolution equations for pressure and enthalpy of the $j$-th particle derived in the last section can now be specified for ideal hydrodynamics.
	The interacting terms of the equations of motion $ \vec E_{\mathrm I_{p,j}}$ and $  E_{\mathrm I_{\mathcal H,j}}$ can be read off Eq.~\eqref{SPH_interaction_operator_ideal}.

    The term on the right-hand side in the evolution equation for the velocity-density Eq.~\eqref{general_int_euler} then reads
	\begin{align}
	- &  \ave{  \tfrac{1}{m}\vec E_{\mathrm I_p}  \, \rho(t, \vec q)} ^{(n-1)}\nonumber \\
	 &=  - \ave{  \frac{2}{5 m \bar \rho}  \sum_{(i,j)} \int \mathrm d^3q_1 \, \rho_j(\vec q_1)\,  \mathcal H_i  \,  \partial_{q_1} \, \rho_i(\vec q_1) \, \rho_j( \vec q)}^{(n-1)}  \nonumber\\
	 &=  - \ave{  \frac{2}{5 m \bar \rho}  \sum_{(i,j)} \rho_j(\vec q)\,  \mathcal H_i  \,  \partial_{\vec q} \, \rho_i(\vec q) }^{(n-1)}  \nonumber\\
	  &=    - \frac{1}{ m \bar \rho}  \ave{\rho(\vec q)  \,  \partial_{\vec q} \, P(\vec q) }^{(n-1)} \;.
	 \end{align}
    Here, we used the relation between two density operators with the same particle index,
    \begin{align} \label{delta_relation}
    	\left<\int \mathrm d^3 q_1 \,f(\vec q_1) \, \rho_j(\vec q_1) \, \rho_j (\vec q) \right>
    	= \left< f(\vec q) \, \rho_j(\vec q) \right>\;,
    \end{align}
    appreciating the distribution form of the densities and the integrals contained in the expectation value.
 
	The term appearing in the evolution equation for the energy-density can be treated in an analogous way,
	\begin{align}
	&- \ave{ \tfrac{3}{5} E_{\mathrm I_{\mathcal H}}\, \rho(t,\vec q )}^{(n-1) } \\
	&= - \frac{1}{\bar \rho} \left(\ave{  \rho(\vec q) \, \partial_{\vec q}  \, P\vec u (\vec q) }^{(n-1) }- \ave{  \rho \vec u (\vec q) \,\partial_{\vec q}  \, P(\vec q) }^{(n-1) } \right) \nonumber\;.
	\end{align}

	\subsection{Diffusive terms}\label{mac_evol_diffusive_terms}

	For viscous hydrodynamics, the diffusive parts of the equations of motion Eq.~\eqref{SPH_momentum_change_viscous} and Eq.~\eqref{SPH_enthalpy_change_viscous} must be included in addition to the interactions of ideal hydrodynamics. In the momentum conservation equation, these lead to the additional term
	\begin{align}
	- &\ave{  \tfrac{1}{m} E_{\mathrm I_p}^a  \, \rho(t, \vec q)} ^{(n-1)}
	 =\frac{c_{ab}}{\bar \rho} \ave{ \rho(\vec q)\,   \partial_{ q}^b \, \left( \partial_{q}^a \, \rho u^b (\vec q) \right)_\text{s}} ^{(n-1)}\,,
	\end{align}
		where the factor ${1}/{m}$ was used to convert the appearing momentum to a velocity.
	Assuming locally isotropic and homogeneous diffusion, the diffusive constants $c_{ab}$ take the form defined in Eq.~\eqref{diffusion_constants}. The terms then reduce to
		\begin{align}
		- \ave{  \tfrac{1}{m}\vec E_{\mathrm I_p} \, \rho(t, \vec q)} ^{(n-1)} & =\frac{\nu}{ \bar \rho} \ave{  \rho(\vec q)\,  \partial_{\vec q}^2 \, \rho \vec u(\vec q) } ^{(n-1)} \nonumber \\
		&+ \frac{\zeta+ \tfrac{\nu}{3}}{\bar \rho}  \ave{ \rho(\vec q) \, \partial_{\vec q}  \left(\partial_{\vec q} \cdot \rho \vec u(\vec q)\right) } ^{(n-1)} \;.
		\end{align}
	The additional term in the energy evolution equation is
	\begin{align}
	&- \ave{ \tfrac{3}{5} E_{\mathrm I_{\mathcal H}}\, \rho(t,\vec q )}^{(n-1) } \\
	&= \frac{\mu}{\bar \rho} \ave{ \tfrac{3}{5} \sum_{(i,j)} \int \mathrm d^3 q_1 \nonumber
		\, \rho_j(\vec q_1) \,  \mathcal H_i   \, \partial_{q_1}^2\, \rho_i(\vec q_1) \, \rho_j(\vec q )}^{(n-1) } \\
	&+ \frac{c_{ab}}{\bar \rho}  \left< \sum_{(i,j)} \int \mathrm d^3 q_1\, \rho_j(\vec q_1)\, 	\left[  u_i^b - u^b_j\right]\left(p_i^b \,  \partial_{q_1}^a\, \right)_\text{s}\partial_{q_1}^a \,\rho_i(\vec q_1) \right> ^{(n-1) }.\nonumber
	\end{align}
	Using the expression for the diffusive stress energy tensor Eq.~\eqref{stress_energy_tensor} we can simplify the term to

	\begin{align}
		- \ave{ \tfrac{3}{5} E_{\mathrm I,{\mathcal H}}\, \rho(t,\vec q )}^{(n-1) }
		%
		& = \frac{\mu}{\bar \rho} \ave{
		 	\rho(\vec q) \, \partial_{\vec q}^2\, \epsilon(\vec q) }^{(n-1) } \\
		 & + \frac{1}{\bar \rho}  \left<\rho(\vec q) \, \partial_{ q}^a \, T^{ab} u^b (\vec q) \right> ^{(n-1) }  \\
		 &- \frac{1}{\bar \rho}  \ave{\, \rho u^b(\vec q)\, \partial_{q}^a \, T^{ab}(\vec q) } ^{(n-1) } \;. \nonumber
	\end{align}

	\subsection{Full equations of the mesoscopic model}

	Collecting all the above terms, we arrive at the full macroscopic equations. On their left-hand sides, we collect all terms belonging to ideal hydrodynamics, while the terms arising from diffusive effects in viscous hydrodynamics appear on the right-hand side.
	\begin{widetext}
		\begin{align} \label{macroscopic_hydro_equ_continuity}
		&\partial_t \ave{\rho(t, \vec q )}^{(n)}  + \, \partial_{\vec q} \cdot\ave{\rho\vec u (t,\vec q)}^{(n)} = 0\\
		\label{macroscopic_hydro_equ_NS}
		&\partial_t \ave{\rho \vec u (t, \vec q)}^{(n)}  +\, \partial_{\vec q}\cdot \ave{\rho\vec u \otimes \vec u(t,\vec q)}^{(n)}  + \frac{1}{ m \bar \rho}  \ave{\rho(\vec q)  \,  \partial_{\vec q} \, P(\vec q) } ^{(n-1)}  =\frac{\nu}{ \bar \rho} \ave{  \rho(\vec q)\,  \partial_{\vec q}^2 \, \rho \vec u(\vec q) } ^{(n-1)} + \frac{\zeta+ \tfrac{\nu}{3}}{\bar \rho}  \ave{ \rho(\vec q) \, \partial_{\vec q}  \left(\partial_{\vec q} \cdot \rho \vec u(\vec q)\right) } ^{(n-1)}   \\
		\label{macroscopic_hydro_equ_energy}
		&\partial_t \ave{\varepsilon(t, \vec q)}^{(n)}  + \,\partial_{\vec q} \cdot\ave{\varepsilon\vec u (t, \vec q) }^{(n)}+\frac{1}{\bar \rho} \left(\ave{  \rho(\vec q) \, \partial_{\vec q}  \, P\vec u (\vec q) }^{(n-1) }- \ave{  \rho \vec u (\vec q) \,\partial_{\vec q}  \, P(\vec q) }^{(n-1) } \right)
		= \frac{\mu}{\bar \rho} \ave{
			\rho(\vec q) \, \partial_{\vec q}^2\, \epsilon(\vec q) }^{(n-1) } \\
		& \hspace{9.5cm}+ \frac{1}{\bar \rho}  \left( \left<\rho(\vec q) \, \partial_{ q}^a \, T^{ab} u^b (\vec q) \right> ^{(n-1) } 
		-   \ave{\, \rho u^b(\vec q)\, \partial_{q}^a \, T^{ab}(\vec q) } ^{(n-1) }  \right)\;. \nonumber
		\end{align}
	\end{widetext}

	Apart from a few subtle differences that we will discuss in the following, these equations match the equations of conventional hydrodynamics.

	One slight difference is due to the particle nature of KFT which shows most clearly in the energy conservation equation: In the conventional equation appears a derivative acting on a velocity-field, which cannot be expressed in KFT. Instead, this term is replaced by the difference of two terms on the left-hand side of the equation which together describe the same quantity.

	Also, all non-convective terms differ in comparison to the conventional terms by a factor of ${\rho(x)}/{\bar \rho}$. This is caused by the approximation $\rho^{-1}(x) \approx \bar \rho^{-1}$ that was necessary for the construction of the interaction operator. The interpretation of this approximation becomes clearer with the following consideration: Let us assume we could divide the equations by a density (which is formally not possible in KFT). Then the only difference to the conventional fluid equations would be the replacement of the inverse density by an inverse mean density. And as the inverse density captures the inertia of the fluid, this approximation neglects fluctuations in the inertia. We believe that this approximation is acceptable in most applications. If this is not the case, interactions approximating the inverse density at higher orders can be constructed.

	Finally, all terms show the traces of the truncation after a certain perturbation order in the interaction. In particular, the non-convective terms are reduced by exactly one order in the interaction. Hence, they vanish in calculations in the free theory  ($n=0$) but appear already at first perturbation order ($n=1$). Thus, we believe that the KFT fluid model is a good starting point to access statistical properties of fluid ensembles. To reach into strongly non-linear regimes we expect that a resummation of the interaction terms, for example in the style of \cite{2018arXiv180906942L}, will be necessary that we plan to present in a future paper.

\clearpage

\begin{widetext}
\section{Exemplary calculation: The onset of dynamics in an over-density} \label{example-calculation}

	To illustrate the behaviour of the KFT fluid model, we present the onset of dynamics in a toy ensemble in which expectation values of the relevant macroscopic fields can be calculated analytically. We choose an ensemble of initially uncorrelated fluid particles with zero initial velocity and constant enthalpy $\mathcal{H}_0$ per particle. They are distributed in space according to a radially decreasing density distribution with Gaussian profile. As pressure is proportional to energy-density, this corresponds to a high pressure in the centre that drops off radially. The pressure gradient is the starting point for dynamics. For simplicity, we just take the interactions of ideal hydrodynamics into account.

	The initial distribution of this ensemble reads
	\begin{align}
	P[\tens q^{(\mathrm i)}, \tens p^{(\mathrm i)},  \mathcal H  \hspace{-3.5mm} \mathcal H ^{(\mathrm i)}] = \prod_{j=1}^N \dirac(\vec p_j^{\,(\mathrm i)}) \dirac(\mathcal H_j^{\,(\mathrm i)}- \mathcal H_0)  \frac{1}{\left(2 \pi \sigma^2\right)^{\frac{3}{2}}}\exp\left(-\frac{(\vec q_j^{\,(\mathrm i)})^2}{2 \sigma ^2}\right)\,.
	\end{align}

	We imagine the Gaussian density profile on top of a  homogeneous background density $\bar \rho$ that enters into the interaction operator. Apart from that, the background does not influence the dynamics. 

	For the calculation, the interaction operator Eq.~\eqref{SPH_interaction_operator_ideal} is best written in its Fourier representation (since the derivative turns into a multiplication with a wave vector).

	\begin{align}
	S_I^\text{\tiny id-hydro} =  \frac{2}{5 \bar \rho}  \sum_{(i,j)} \int \frac{\mathrm d1}{(2\pi)^3}\, (\mathrm i \vec k_1) \hat{\rho}_j(-1) \hat{\mathcal H}_i  \hat{\rho}_i(1)\,  \hat \chi_{p_j} (1)  +  \frac{2}{3 \bar \rho} \, \sum_{(i,j)} \int \frac{\mathrm d1}{(2 \pi)^3} \, (\mathrm i \vec k_1) \,\hat{\rho}_j(-1) \left[  \hat{\vec u}_i - \hat{\vec u}_j\right] \,\hat{\mathcal H}_i  \hat{ \rho}_i(1) \, \hat \chi_{\mathcal H_j} (1)\,,
	\end{align}
	where now $\mathrm d 1 = \mathrm d^3 k_1 \mathrm d t_1$.

	For the ensemble we calculate the time-dependent expectation values for the density, velocity-density and enthalpy-density
	\begin{align}\label{eq:interaction-op-fourier}
	\ave{\rho(\vec k, t)}^{(n)} &=\frac{\ii^n}{n!} \sum_j  \hat \rho_j(\vec k,t) \left(\hat S_\mathrm I^\text{\tiny id}\right)^n \tilde Z_0\left[  \tens J, \tens K\right]\bigg|_{\tens J = 0= \tens K}\\
	\ave{\rho\vec u(\vec k, t)}^{(n)}& =\frac{\ii^n}{n!} \sum_j  \hat \rho_j(\vec k,t) \hat u_j(t) \left(\hat S_\mathrm I^\text{\tiny id}\right)^n \tilde Z_0\left[  \tens J, \tens K\right]\bigg|_{ \tens J = 0= \tens K} \\
	\ave{\rho\mathcal H(\vec k, t)}^{(n)}& =\frac{\ii^n}{n!} \sum_j  \hat \rho_j(\vec k,t) \hat{\mathcal H}_j(t) \left(\hat S_\mathrm I^\text{\tiny id}\right)^n \tilde Z_0\left[  \tens J, \tens K\right]\bigg|_{ \tens J = 0= \tens K}\;.
	\end{align}
	up to third order in the interactions. $S_\mathrm I$ is the interaction of ideal hydrodynamics as in Eq.~\eqref{eq:interaction-op-fourier} and $\tilde Z_0\left[  \tens J, \tens K\right]$ is the free generating functional for hydrodynamics as in Eq.~\eqref{mesoscopic_generating_free_functional}.

	The derivation could in principle be performed in a brute force manner. However, there are several ways to significantly shorten the calculations. By establishing a time-ordering, using the notion of `contractions' introduced in Appendix~\ref{appendix_macroscopic_evolution}, and taking the form of the initial conditions into account, many of the terms can be seen to vanish right away.

	Here we sketch the simplified calculation for the second order correction of the enthalpy-density as it shows all important features and methods of performing the calculation. We use a strongly simplified notation that shows only the very basic form of the interaction operator and drops all prefactors. We get
	\begin{align}
	\ave{ \hat{\mathcal{H}} \rho(0)}^{(2)}&\propto \hat S_I^2\,  \hat{\rho {\mathcal H}}(\vec k_0, t_0)\, \tilde Z_0\left[  \tens J, \tens K\right]\bigg|_{ \tens J = 0= \tens K}\\
	&= \sum_\text{\tiny particles}\int \mathrm d2 \, \mathrm d1 
	\left(\hat{\vec E}_p(2) \cdot \hat{\vec{\chi}}_{p,j_2}(2) + \hat E_{\mathcal H}(2) \,\hat \chi_{\mathcal H,j_2} (2) \right)
	\left(\hat{\vec E}_p(1) \cdot \hat{\vec{\chi}}_{p,j_1}(1) + \hat E_{\mathcal H}(1) \,\hat \chi_{\mathcal H,j_1} (1) \right)\,
	\hat {\mathcal H}_{i_0}\,  \hat\rho_{i_0}(0)\, \tilde Z_0\left[  \tens J, \tens K\right]\bigg|_{ 0} \nonumber\;,
	\end{align}
	where the sum runs over all appearing particle indices $i_2$, $j_2 \neq i_2$, $i_1$, $j_1 \neq i_1$, $i_0$. Without loss of generality we demand $t_2<t_1<t_0$ such that in the above ordering all operators may only act to the right in order to guarantee a causal structure. 

	Once the source fields $J$ and $K$ are set to zero only those terms survive in which each $\hat {\vec \chi}_p$ operator is either paired with a density operator or a velocity operator and each $\hat { \chi}_\mathcal H$ is paired with an enthalpy operator. That is, these pairs act on the generating functional in a way that they are non-vanishing once the source fields are set to zero which is not the case for unpaired $\hat {\vec \chi}_p$ operators. This pairing we call `contraction' and denote it by a line connecting the operators. A more detailed description of the contractions' derivation and their exact form is given in Appendix \ref{appendix_macroscopic_evolution}. 

	Specifying the contractions for the $\chi$-operators in the $t_1$ bracket yields two surviving terms
	\begin{align}
	\ave{ \mathcal H\rho(0)}^{(2)}&\propto  \sum_\text{\tiny particles}\int \mathrm d2 \, \mathrm d1 
	\left(\hat{\vec E}_p(2) \cdot \hat{\vec{\chi}}_{p,j_2}(2) + \hat E_{\mathcal H}(2) \,\hat \chi_{\mathcal H,j_2} (2) \right)
	\hat{\vec E}_p(1) \cdot \wick{\c2{\hat{\vec{\chi}}}_{p,j_1}(1) 
		\, \hat {\mathcal H}_{i_0}\, \c2{\hat\rho}_{i_0}(0)}\, \tilde Z_0\left[  \tens J, \tens K\right]\bigg|_{ 0} \\
	&+ \sum_\text{\tiny particles}\int \mathrm d2 \, \mathrm d1 
	\left(\hat{\vec E}_p(2) \cdot \hat{\vec{\chi}}_{p,j_2}(2) + \hat E_{\mathcal H}(2) \, \hat \chi_{\mathcal H,j_2} (2) \right)
	\hat E_{\mathcal H}(1)\, \wick{\c2{\hat \chi}_{\mathcal H,j_1} (1) \, \c2{\hat {\mathcal H}}_{i_0}}
	\,  \hat\rho_{i_0}(0)\,  \tilde Z_0\left[  \tens J, \tens K\right]\bigg|_{ 0} \nonumber\;.
	\end{align}
	For both terms the contractions of the remaining bracket can now be specified. At this point the initial conditions enter. These are chosen such that all initial particle momenta are zero. Hence, after applying all operators, any term that is multiplied by an initial momentum will vanish once the integration over initial conditions is performed. 

	The equation of motion for the enthalpy $\hat E_{\mathcal H}$ is linear in the velocity operator, and the velocity operator extracts a Green's function multiplied with the initial momentum. Hence, it produces a term linear in the initial momentum. The contraction of $\hat{\vec{\chi}}_p$ with the velocity operator removes this linearity such that terms containing this contraction for all $\hat E_{\mathcal H}$ are non-zero,
	\begin{align}
	\ave{ \mathcal H \rho(0)}^{(2)}&\propto  \sum_\text{\tiny particles} \int \mathrm d2 \, \mathrm d1 
	\hat{\vec E}_{p}(2) \cdot \wick{\c3{\hat{\vec{\chi}}}_{p,j_2}(2) \, 
		{\hat{\vec E}}_p(1) \cdot \c2{\hat{\vec{\chi}}}_{p,j_1}(1)  \,\hat {\mathcal H}_{i_0}
		\,\c2{{\hat\rho}}\hspace{-0.5mm}\c3{\vphantom{{\hat\rho}}}_{i_0}(0)}\hspace{+0.4mm}\, \tilde Z_0\left[  \tens J, \tens K\right]\bigg|_{ 0}\\
	&+  \sum_\text{\tiny particles} \int \mathrm d2 \, \mathrm d1 
	\hat{\vec E}_p(2) \cdot \wick{\c2{\hat{\vec{\chi}}}_{p,j_2}(2) \, 
		\c2{\hat{\vec E}}_p(1) }\cdot \wick{\c2{\hat{\vec{\chi}}}_{p,j_1}(1) \, \hat {\mathcal H}_{i_0}
		\,  \c2{\hat\rho}}_{i_0}(0)\, \tilde Z_0\left[  \tens J, \tens K\right]\bigg|_{ 0}\nonumber \\
	&+ \sum_\text{\tiny particles} \int \mathrm d2 \, \mathrm d1 
	\hat{\vec E}_p(2) \cdot \wick{\c2{\hat{\vec{\chi}}}_{p,j_2}(2) \,
		\c2{\hat E}_{\mathcal H}(1)}\, \wick{\c2{\hat \chi}_{\mathcal H,j_1} (1) \,\c2{\hat {\mathcal H}}_{i_0}
		\,  \hat\rho_{i_0}(0) }\, \tilde Z_0\left[  \tens J, \tens K\right]\bigg|_{ 0} \;,\nonumber
	\end{align}
	with 
	\begin{align}
	\wick{\c2{\hat{\vec{\chi}}}_{p,j_2}(2) \,    \c2{\hat{\vec E}}_p(1) } &= \wick{\c2{\hat{\vec{\chi}}}_{p,{j_2}}(2) \, (\mathrm i \vec k_1)\, \c2{\hat{\rho}}_{j_1}(-1)\, \hat{\mathcal H}_{i_1} \, \hat{\rho}_{i_1}(1)} +  \wick{\c2{\hat{\vec{\chi}}}_{p,j_2}(2) \, (\mathrm i \vec k_1) \,{\hat{\rho}}_{j_1}(-1) \, \hat{\mathcal H}_{i_1}  \,\c2{\hat{\rho}}_{i_1}(1)} \;,\\
	\wick{\c2{\,\hat{\vec{\chi}}}_{p,{j_2}}(2)
		\c2{\hat E}_{\mathcal H}(1)} &= \wick{\c2{\hat{\vec{\chi}}}_{p,{j_2}}(2) \,(\mathrm i \vec k_1) \,\hat{\rho}_{j_1}(-1) \,  \c2{\hat{\vec u}}_{i_1}\,\hat{\mathcal H}_{i_1} \, \hat{ \rho}_{i_1}(1)}- \wick{\c2{\hat{\vec{\chi}}}_{p,{j_2}}(2) \,(\mathrm i \vec k_1) \,\hat{\rho}_{j_1}(-1) \,  \c2{\hat{\vec u}}_{j_1}\,\hat{\mathcal H}_{i_1} \, \hat{ \rho}_{i_1}(1)}\;.
	\end{align}
	In total five non-vanishing terms remain which we have to evaluate. For the chosen set up this means performing several Gauss-integrals which we will spare the reader from.

	Our results for the perturbation terms in configuration space are shown in the following table. Here, we used the dimension-less time $\tau$ and radial coordinate $\xi$ 
	\begin{align}
	\tau = t \sqrt{\frac{\mathcal H_0 }{m \sigma^2}}\,, \qquad \xi = \frac{r}{\sigma}\,,
	\end{align}
	the characteristic density and velocity of the ensemble 
	\begin{align}
	\rho_0 = \frac{N}{(2 \pi \sigma^2)^{3/2}}\,, \qquad v_0 = \frac{\sigma}{\tau}\,,
	\end{align}
	and the approximation $N \approx N-1\approx N-2 \approx N-3$ for large particle numbers.

	\begin{table}[h!]
		\setlength\tabcolsep{8pt}
		\setlength\extrarowheight{12pt}
		\begin{tabular}{@{\extracolsep{\fill}} |c||c|c|c|}
			\hline  {order} & density & velocity-density & enthalpy-density\\
			\hline \hline  {0\textsuperscript{th}} & $\ave{\rho(\xi,\tau)}^{(0)}=  \rho_0\, \e^{-\frac{1}{2 }\xi^2} $& $\ave{\vec u \rho(\xi,\tau)}^{(0)}= 0$  & $\ave{\mathcal H \rho(\xi,\tau)}^{(0)}=\mathcal{H}_0 \ave{\rho (r, t)}^{(0)}$ \\
			\hline {1\textsuperscript{st}} & $\ave{\rho(\xi,\tau)}^{(1)}=\rho_0 \frac{\e^{-\xi^2} \tau^2 \left(2\xi^2-3\right)}{5} \frac{\rho_0}{\bar\rho}$  &$ \ave{\vec u\rho(\xi,\tau)}^{(1)}=\rho_0 v_0 \,\frac{ \e^{-\xi^2} \, \tau\,   2 \xi }{5}  \frac{\rho_0}{\bar \rho}$ & $\ave{\mathcal H \rho(\xi,\tau)}^{(1)}=\mathcal{H}_0 \ave{\rho (r, t)}^{(1)}$ \\
			\hline {2\textsuperscript{nd}} & $\ave{\rho(\xi,\tau)}^{(2)}=\rho_0 \frac{ \e^{-\frac{3}{ 2 }\xi^2}\tau^4\left(2\xi^4-68\xi^2+ 33 \right)}{50} \left(\frac{\rho_0}{\bar\rho}\right)^2$ & $\ave{\vec u\rho(\xi,\tau)}^{(2)}=\rho_0 v_0 \,
			\frac{\exp^{-\frac{3}{2} \xi^2} \tau^3\, \left(32\xi^3-44 \xi \right)}{75} \, \left(\frac{\rho_0 }{\bar \rho}\right)^2$
			&\shortstack{ $ \ave{\mathcal H \rho(\xi,\tau)}^{(2)}=\mathcal{H}_0 \ave{\rho (r, t)}^{(2)}    $ \\
				$+ \mathcal{H}_0 \rho_0  \frac{\e^{-\frac{3 }{2} \xi^2} \tau^2 \left(2 \xi^2- 6 \right) }{15}\left(\frac{\rho_0}{\bar \rho}\right)^2$ } \\
			\hline {3\textsuperscript{rd}} & \shortstack{ $\ave{\rho(\xi,\tau)}^{(3)}=\rho_0 \frac{\e^{-2\xi^2}\tau^4 \left(16\xi^4+ 68\xi^2+ 36\right)}{225} \left(\frac{\rho_0}{\bar \rho}\right)^3$ \\
				$ + \rho_0 \frac{\e^{-2 \xi^2} \tau^6 \left(1384\xi^6+ 7006\xi^4+8394\xi^2 + 1998\right)}{1875} \left(\frac{\rho_0}{\bar \rho} \right)^3$ }& \shortstack{$\ave{\vec u\rho(\xi,\tau)}^{(3)}=\rho_0 v_0  \frac{\e^{-2\xi^2} \tau^3 \left(16\xi^3+48\xi\right) }{225} \left(\frac{\rho_0}{\bar \rho}\right)^3$\\
				$+ \rho_0 v_0 \,  \frac{\e^{-2 \xi^2} \tau^5 \left(
					692\xi^5+ 2292\xi^3+ 1332\xi\right)}{625} \left(\frac{\rho_0}{\bar \rho}\right)^3 $} &  \shortstack{$\ave{\mathcal H \rho(\xi,\tau)}^{(3)}= \mathcal{H}_0 \ave{\rho (r, t)}^{(3)}$ \\
				$+ \mathcal{H}_0 \rho_0  \frac{\e^{-2 \xi^2}\tau^4\left(91 \xi^4 -389 \xi^2 + 264 \right)}{225}  \left(\frac{\rho_0}{\bar\rho}\right)^3$}\\
			\hline
		\end{tabular}
	\end{table}

	Figure \ref{fig:all-plots} visualises these results. It shows the density, velocity, enthalpy-density (that is proportional to the pressure) and the enthalpy of the ensemble in zeroth order as well as their first three corrections. While the density and enthalpy-density are expectation values by themselves, the velocity and enthalpy must be extracted from the combination of several expectation values. The $n$-th order velocity and enthalpy corrections are
	\begin{align}
	\vec v^{(n)}(0) = \frac{\sum_{m=0}^n \ave{\vec u \rho (0)}^{(m)} }{\sum_{m=0}^n \ave{ \rho (0)}^{(m)} }- \frac{\sum_{m=0}^{n-1} \ave{\vec u \rho (0)}^{(m)} }{\sum_{m=0}^{n-1} \ave{ \rho (0)}^{(m)} }\;, \qquad     
	\mathcal H^{(n)}(0) = \frac{\sum_{m=0}^n \ave{\mathcal H \rho (0)}^{(m)} }{\sum_{m=0}^n \ave{ \rho (0)}^{(m)} }- \frac{\sum_{m=0}^{n-1} \ave{\mathcal H \rho (0)}^{(m)} }{\sum_{m=0}^{n-1} \ave{ \rho (0)}^{(m)} }\;.
	\end{align}

	In the free theory, the ensemble is static due to the zero initial velocities. Only once the interactions are turned on the dynamics set in: The over-density combined with a constant enthalpy causes a pressure gradient which drives particles away from the centre. The onset and development of the particles' velocities can be seen clearly already in the first order correction. This motion reduces the density and hence the pressure in the centre. The enthalpy of the particles remains unchanged in the first order correction. Only in the second order correction, the enthalpy is changed by pressure-volume work caused by gradients in the particles' velocities. At the same time, the particle motion gains a more complex structure. The second order correction partially compensates the outward motion of the particles which leads to a lessened reduction of the central density. 

	This nicely shows the cross-talk between the two parts of the interaction operator: Pressure gradients set particles in motion which modifies both position and enthalpy of the particles. This is a modification of the pressure which in turn changes the motion of the particles. In the third order correction, even more complex behaviour arises that is hard to explain intuitively. 

	Taking a look at amplitudes, density corrections decrease about one order of magnitude with each higher order correction. Amplitudes of enthalpy density corrections also decrease quickly with higher orders, but this is not true for the velocity and the enthalpy. The latter two are the quantities that are not directly derivable in KFT but must be formed as a ratio between velocity density or enthalpy density and density. While all directly derivable quantities fall off quickly, the combined ones do not necessarily. 

	Finally, we would like to stress that this example was chosen to help with an intuitive understanding of the KFT fluid model. However, it does not reflect the strength of KFT which lies in the calculation of higher-order correlation functions, especially in statistically homogeneous ensembles that start from correlated initial conditions. 

	\clearpage

	\begin{figure}[H]
		\centering
		\includegraphics[width=\textwidth]{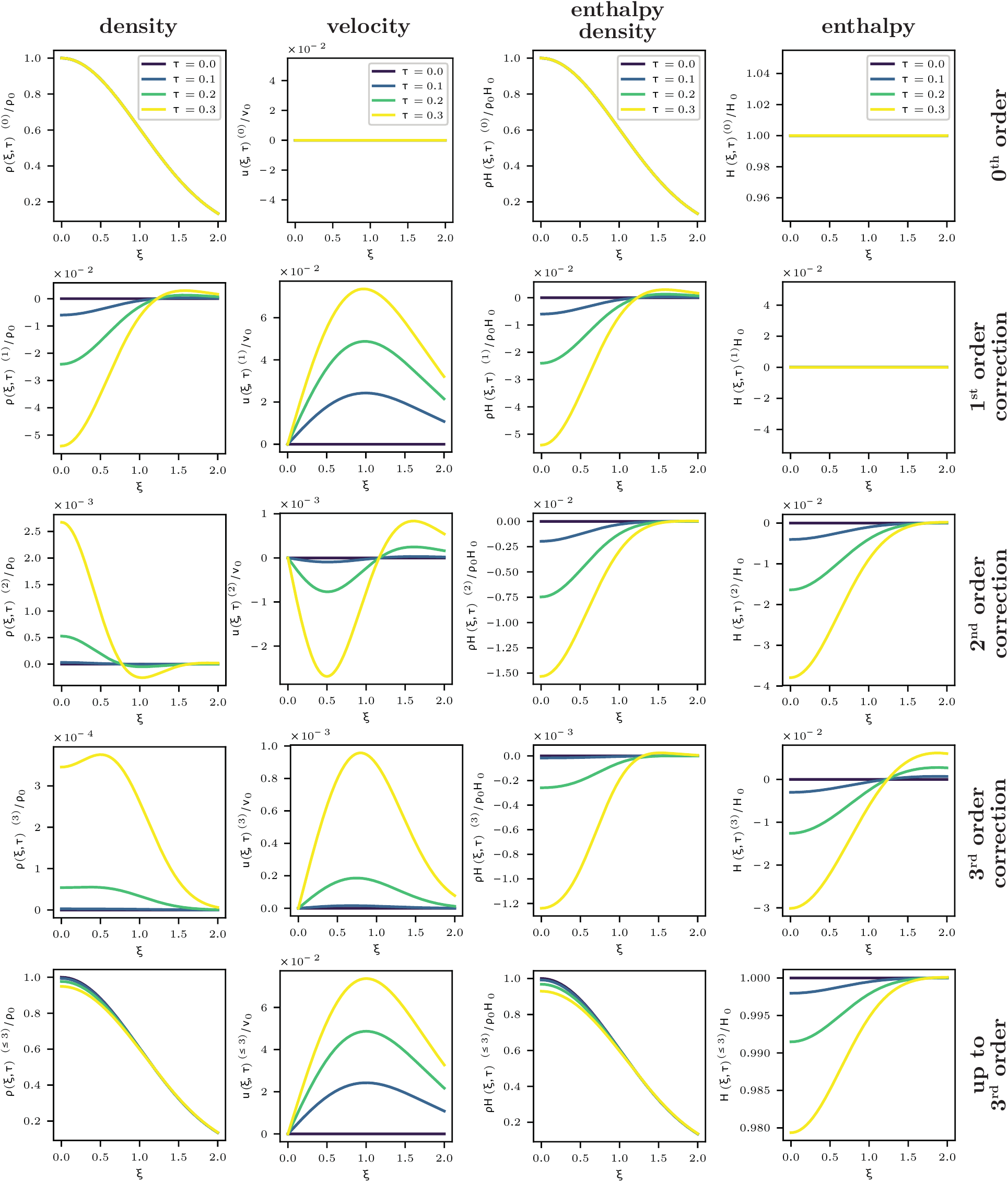}
		\caption{\label{fig:plots}
			Four macroscopic quantities are plotted here from left to right: density, velocity, enthalpy-density (proportional to pressure), and enthalpy. From top to bottom we plotted the 0th order, the $1$st order correction, the $2$nd order correction, the $3$rd order correction, and in the last line sum of the $0$th order and all corrections. The different colours encode the time evolution. For simplicity, we chose ${\rho_0}/{\bar \rho}=1$. \newline At $0$th order the ensemble remains static due to the choice of the initial conditions. In the first order correction, the pressure gradient accelerates particles. This also leads to a decrease of the central density. In the second and third order the divergent velocities in addition lead to a change in the particle's enthalpy and more complex structures in both velocity and density appear.}
			\label{fig:all-plots}
	\end{figure}
	
	\end{widetext}

\section{Summary and Conclusion}

	In this work, we set up an effective model including hydrodynamics into the framework of KFT. The model is based on the introduction of mesoscopic particles similar to fluid elements in conventional hydrodynamics. Next to their position and momentum, each of these particles is equipped with an enthalpy (or possibly a local copy of the stress-energy tensor).

	For these particles, the free generating functional was constructed in Section~\ref{free_theory_mesoscopic_particles}. We complemented the free theory by appropriate interaction operators that we derived in analogy to smoothed particle hydrodynamics. The final operators are Eq.~\eqref{SPH_interaction_operator_ideal} for ideal hydrodynamics and Eq.~\eqref{SPH_interaction_operator_viscous} for viscous effects.

	For a generalised interaction in the mesoscopic model, we derived macroscopic evolution equations for the density, velocity-density and energy  Eq.~\eqref{general_int_continuity_f} to Eq.~\eqref{general_int_energy_f} that we specified to the hydrodynamic model in Eq.~\eqref{macroscopic_hydro_equ_continuity} to Eq.~\eqref{macroscopic_hydro_equ_energy}.

	Already at first order perturbations in the interaction, these evolution equations take the form of the continuity equation, Euler/Navier-Stokes equation, and energy conservation equation of conventional hydrodynamics. The most physically significant difference lies in a necessary approximation: Fluctuations in the inertia of the system can only be taken into account perturbatively.

	We also presented an exemplary calculation within the KFT hydrodynamic model which visualises the onset of dynamics in increasing orders in perturbation theory. These dynamics already exhibit characteristic non-linear, fluid-like behaviour. However, the example also shows that a resummation of the interaction is necessary. A KFT-resummation was already performed in \cite{2018arXiv180906942L} and will be applied to the fluid model in a future paper.

\begin{acknowledgements}
	We want to thank Björn Malte Schäfer, Manfred Salmhofer, Jürgen Berges and Robert Reischke for many helpful discussions. 
	This work was supported in part by the Heidelberg Graduate School of physics, the Collaborative Research Centre TR 33 "The Dark Universe" of the German Science Foundation and the Heidelberg Center for Quantum Dynamics. 

\end{acknowledgements}

\appendix

\section{Macroscopic evolution equations for a general interaction}
    \label{appendix_macroscopic_evolution}
    In this appendix we present the derivation of macroscopic evolution equations for the density, velocity-density, and energy-density that arise from the generating functional
    \begin{align}
    \tilde Z[H,\tens J,\tens K] = \e^{\ii \hat S_{\mathrm I}} \, \tilde Z_0[H,\tens J ,\tens K]
    \end{align}
    with the general interaction operator
    \begin{align}
    \ii \hat S_{\mathrm I} =& \ii \int \d t  \sum_j \hat{\vec E}_{\mathrm I_p}(\vec q_j, \vec p_j, {\mathcal H}_j) \cdot \hat{\vec \chi}_{p_j}\\
    \nonumber &+ \ii \int \d t  \sum_j \hat E_{\mathrm I_{\mathcal H}}(\vec q_j, \vec p_j, \mathcal H_j) \, \hat \chi_{{\mathcal H}_j} \;. \label{general_interaction_fluid_appendix}
    \end{align}
    The interaction operator consists of the yet unspecified operator expressions of the interaction part of the equations of motion for momentum and enthalpy
    $\hat E_{\mathrm I_ p}(\vec q_j, \vec p_j, \mathcal H_j)$ and $\hat E_{\mathrm I_{\mathcal H}}(\vec q_j, \vec p_j, \mathcal H_j)$ as well as the operators $\hat{\vec \chi}_{p_j}$ and $\hat \chi_{{\mathcal H}_j}$ which correspond to a functional derivative with respect to $K_{p_j}$ and $K_{{\mathcal H}_j}$, respectively.
    
    We start the calculation by taking a looking at a $\hat{\vec \chi}_{p_j}$ operator acting on the free functional
    \begin{align}
    \hat{\vec \chi}_{p_j} \, \tilde Z_0[H,\tens J &,\tens K] \big|_0= \der{}{K_{p_j}} \, \tilde Z_0[H,\tens J ,\tens K]\big|_0
    \\ &=  \der{}{K_{p_j}} \, \int \d\Gamma_{\mathrm i}\, \nexp{\ii \int_{\mathrm i}^{\mathrm f} \d t \, \langle\tens J (t), \bar{\boldsymbol \varphi}(t)\rangle }\bigg|_0\nonumber\\
    & =  \int \d\Gamma_{\mathrm i}\, \left<\tens J (t), \frac{\delta \bar{\boldsymbol \varphi}(t)}{\delta K_{p_j}}\right>\, \nexp{\ii \int_{\mathrm i}^{\mathrm f} \d t \, \langle\tens J (t), \bar{\boldsymbol \varphi}(t)\rangle } \bigg|_0\nonumber \,.
    \end{align}
    
    This term is linear in $\mathbf{J}$ and will vanish once the source fields are set to zero at the end of each calculation. The only way for this term not to vanish is the application of another collective field operator containing at least one $J$-derivative such that the overall term does not depend on $\mathbf{J}$. The same consideration holds for the $\hat{\vec \chi}_{\mathcal H}$-operator. 
    
    Hence, the $\hat{\vec \chi}$-operators only yield non-zero terms if they appear in combination with another operator. We mark the couple of operators by a connecting line and shall call this a `contraction', for example
    \begin{align}
    \wick{\c2{\hat \chi}_{p_j} (t') \, \c2{\hat\rho}_i(t, \vec k)\,\tilde  Z_0 [H, \tens{J}, \tens{K}]}\big|_0\,.
    \end{align}
    
    Any field expectation value in first or higher order perturbation theory reduces to the sum over all possible contractions which fulfil an established time ordering.
    
    As the functional derivatives commute, the order of the operators within a contraction is arbitrary. However, calculations turn out to be much more convenient if the collective field operator is applied before the $K$-derivatives. For example, the contraction of a density operator and a $\hat \chi_{p_j}$-operator reads
    \begin{align}
    \wick{\c2{\hat \chi}_{p_j}(t')  \, \c2{\hat\rho}_i(t, \vec k)\,\tilde  Z_0 [H, \tens{J}, \tens{K}]}\big|_0 \hspace{-3.2cm}&\\
    &= \der{}{K_{q_j}(t')} \, \nexp{- i \vec k \cdot \der{}{J_{q_i}(t)}} \, \tilde Z_0 [H, \tens{J}, \tens{K}] \big|_0 \nonumber\\
    &= \der{}{K_{q_j}(t')} \, \int \d\Gamma_{\mathrm i}\, \nexp{- i \vec k \cdot \bar q_i(t)} \, \nexp{\ii \int_{\mathrm i}^{\mathrm f}\hspace{-1mm} \d t \, \langle\tens J (t), \bar{\boldsymbol \varphi}(t)\rangle }\nonumber \\
    &=  \vec k^\ntop g_{qp}(t-t')\,\theta(t-t')\,\delta_{ij} \int \d\Gamma_{\mathrm i}\, \nexp{- i \vec k \cdot \bar q_i(t)} \nonumber\\
    &\hspace{4cm}\,\times \, \nexp{\ii \int_{\mathrm i}^{\mathrm f}\hspace{-1mm} \d t \, \langle\tens J (t), \bar{\boldsymbol \varphi}(t)\rangle }\;, \nonumber
    \end{align}
    where the definition of $\bar q_i$ in Eq.~\eqref{hydro_bar_phi} was used. A second term arises from the $K$-derivative acting on the second exponential term. This term does not appear since it would again be linear in $J$ and vanish once the source fields are set to zero.
    
    In a last step we can identify the first exponential with the term arising from a density operator and rewrite the contraction as
    \begin{align}
    \wick{\c2{\hat \chi}_{p_j} (t')  \, \c2{\hat \rho}_i(t, \vec k)\,\tilde  Z_0 [H, \tens{J}, \tens{K}]\big|_0 }\hspace{-2.5cm}&\\
    &=  \vec k^\ntop g_{qp}(t-t')\,\theta(t-t')\,\delta_{ij}  \, \hat\rho_i(t, \vec k)\,\tilde  Z_0 [H, \tens{J}, \tens{K}] \big|_0\nonumber\;.
    \end{align}
    
    All other contractions which will appear in the scope of this work can be derived in a similar manner. Using a shorthand notation which leaves out the free functional, they read
    \begin{align}
    &\wick{\c2{\hat \chi}_{p_j} (t') \, \c2{\hat \rho}_i(t,\vec k) \,= \vec k^\ntop\, g_{qp}(t-t') \,\theta(t-t')\,\delta_{ij} \,\hat\rho_i(t, \vec k) \label{contr_p_rho}\;,}\\
    &\wick{\c2{\hat \chi}_{{\mathcal H}_j} (t') \, \c2{\hat\rho}_i(t,\vec k) = \vec k^\ntop\, g_{q{\mathcal H}}(t-t')\,\theta(t-t') \,\delta_{ij} \,\hat\rho_i(t, \vec k) \equiv 0 \label{contr_T_rho}\;,}\\
    &\wick{\c2{\hat \chi}_{p_j} (t') \,\c2{ \hat {\vec u}}_i(t) =  \frac{\ii}{m} \,g_{pp}(t-t')\,\theta(t-t') \,\delta_{ij} \label{contr_p_vel}} \;,\\
    &\wick{\c2{\hat \chi}_{{\mathcal H}_j} (t') \, \c2{\hat {\vec u}}_i(t) =  \frac{\ii}{m} \,g_{p{\mathcal H}}(t-t')\,\theta(t-t')\, \delta_{ij} \equiv 0 \label{contr_T_vel} \;,}\\
    & \wick{\c2{\hat \chi}_{p_j} (t') \, \c2{\hat{\mathcal H}}_i(t) =  \ii \,g_{{\mathcal H}p}(t-t')\,\theta(t-t') \,\delta_{ij} \equiv 0  \label{contr_p_stress}\;,}\\
    &\wick{\c2{\hat \chi}_{{\mathcal H}_j} (t') \, \c2{\hat{\mathcal H}}_i(t) =  \ii \,g_{{\mathcal H}{\mathcal H}}(t-t')\,\theta(t-t') \,\delta_{ij}} \label{contr_T_stress}\;,
    \end{align}
    where three of the six objects vanish identically due to the definition of the Green's functions in  Eq.~\eqref{hydro_greens_functions}.

    Any contraction involving combined field operators can be traced back to these fundamental contractions, for instance,
    \begin{align}
    \wick{\c3{\hat \chi}_{p_j} (t') \,\c3{ \hat{\rho\vec{u}} }_i(t,\vec k)}= &\, \wick{\c2{\hat \chi}_{p_j} (t') \, \c2{\hat \rho}_i(t,\vec k)\,\hat {\vec u}_i(t) }
    + \wick{\,\c2{\hat \chi}_{p_j} (t') \, \hat \rho_i(t,\vec k)\,\c2{\hat {\vec u}}_i(t)}\;.
    \end{align}
    
    To derive the macroscopic evolution equations, we will need time derivatives of both the field operators and all non-vanishing contractions. Still using the shorthand notation which leaves out the free functional, we obtain for the time derivatives of the field operators
    \begin{align}\label{tder_density_field}
    &\partial_t \,\hat \rho_j(t,\vec k)\,\tilde  Z_0 [H, \tens{J}, \tens{K}]\big|_0 = \,\partial_t \, \int \mathrm d \Gamma \,\exp\left\{-\ii \vec k \cdot \left(\vec q_j^{\,(\mathrm i)} + \tfrac{(t-t_\mathrm i)}{m}\vec p_j^{\,(\mathrm i )}\right)\right\}\nonumber \\
    &\phantom{\partial_t \,\o_{\rho_j}(t,\vec k) }
    = -\ii \vec k \cdot  \int \mathrm d \Gamma\,\frac{1}{m} \vec p_j^{\,(\mathrm i )}\,\exp\left\{-\ii \vec k\cdot \left(\vec q_j^{\,(\mathrm i)} + \tfrac{(t-t_\mathrm i)}{m}\vec p_j^{\,(\mathrm i )}\right)\right\} \nonumber\\
    &\phantom{\partial_t \,\o_{\rho_j}(t,\vec k) } = -\ii \vec k \cdot\hat{\vec u}_j(t) \, \hat\rho_j(t,\vec k)\, \tilde  Z_0 [H, \tens{J}, \tens{K}]\big|_0  \;,
    \end{align}
    where $\tfrac{1}{m} \vec p_j^{\,(\mathrm i )}$ was identified with the expression arising from a velocity operator acting on the free functional and
    \begin{align}
    \partial_t\, \hat{\vec u}_j(t)\, \tilde  Z_0 [H, \tens{J}, \tens{K}]\big|_0 &= \partial_t \, \int \mathrm d \Gamma \, \tfrac{1}{m}\vec p_j^{\,(\mathrm i )} = 0 \label{tder_velocity}\;,\\
    \partial_t \, \hat{\mathcal H}_j(t)\, \tilde  Z_0 [H, \tens{J}, \tens{K}]\big|_0 &= \partial_t \,\int \mathrm d \Gamma\, {\mathcal H}_j^{\,(\mathrm i )} = 0 \label{tder_enthalpy}\;.
    \end{align}
    
    Omitting the generating functional in the notation, the time derivatives of the contractions yield
    \begin{align} \label{tder_contr_p_rho}
    \partial_t \, &\left(\,\wick{\c1{\hat \chi}_{p_j} (t') \,\c1{\hat\rho}_i(t,\vec k)}\right) \\
    &  \hspace{5mm}= \,\tfrac{1}{m}\vec k^\ntop g_{pp}(t-t') \,\theta(t-t')\,\delta_{ij} \,\hat \rho_i(t, \vec k)\nonumber \\
    &  \hspace{5mm}\phantom{=}+   \vec k^\ntop g_{qp}(t-t')\,\theta(t-t') \,\delta_{ij} \,\left( -\ii \vec k \cdot {(\hat{\rho\vec{u}})_i}(t, \vec k)\right) \nonumber\\
    & \hspace{5mm}\phantom{=} + \vec k^\ntop g_{qp}(t-t')\,\dirac(t-t') \,\delta_{ij}\, \hat \rho_i(t, \vec k)\nonumber \\
    &   \hspace{5mm}= \,\wick{\c2{\hat \chi}_{p_j} (t') \, \left(-\ii \vec k \cdot \c2{ (\hat{\rho\vec{u}})}_i(t,\vec k)\right)} \nonumber\\
    & \hspace{5mm}\phantom{=}+ \vec k^\ntop g_{qp}(t-t')\,\dirac(t-t') \,\delta_{ij}\, \hat \rho_i(t, \vec k) \nonumber \,, \\
    \partial_t \, &\left(\,\wick{\c2{\hat \chi}_{p_j} (t') \, \c2{\hat {\vec u}_i}(t,\vec k)}\right) \,\,\,\,=  \frac{\ii}{m} g_{pp}(t-t')\,\dirac(t-t')\, \delta_{ij} \label{tder_contr_p_u} \,, \\
    \partial_t \, &\left(\,\wick{\c2{ \hat{\chi}}_{{\mathcal H}_j} (t') \, \c2{ \hat{\mathcal H} }_i(t,\vec k)}\right) = \, \ii\, g_{{\mathcal H}{\mathcal H}}(t-t')\,\dirac(t-t') \,\delta_{ij} \;,  \label{tder_contr_h_h}
    \end{align}
    where we used that neither $g_{ p p }(t-t')$ nor $g_{\mathcal H \mathcal H}(t-t')$ depend explicitly on time and that $g_{qp}(0) = 0$.
    
    With these time derivatives at hand, we proceed with the evolution equations for the density, velocity-density, and energy-density fields in first order perturbation theory. Starting with the equation for the density field, we get
    \begin{align}
    \partial_t &\ave{\rho(t, \vec k )}^{(1)}   \\
    &= \, \partial_t \, \sum_{\mu,j}\ii \int \d t'\, \vec E_{\mathrm I,p_j}(t')\cdot \wick{\c2{\hat{\vec \chi}}_{p_j}(t')\,\c2{\hat\rho}_\mu(t, \vec k)\,\tilde  Z_0 [H, \tens{J}, \tens{K}]\big|_0 }  \nonumber \\
    &+\, \partial_t\, \sum_{\mu,j}\ii \int \d t'\, E_{\mathrm I,{\mathcal H}_j} \,\wick{\c2{ \hat \chi}_{{\mathcal H}_j}(t')\,\c2{\hat\rho}_\mu}(t, \vec k)\, \tilde  Z_0 [H, \tens{J}, \tens{K}]\big|_0  \nonumber\;.
    \end{align}
    Since the equations of motion themselves do not change in time, the time derivatives act only on the contraction of the $\chi$-operators with the density field. Inserting expressions (\ref{contr_T_rho}) and (\ref{tder_contr_p_rho}) yields
    \begin{align}
    \partial_t & \ave{ \rho(t, \vec k )} ^{(1)} \\
    & = \sum_{\mu,j}\ii \int \d t'\, \vec E_{\mathrm I,p_j}\cdot \wick{\c2{\hat{\vec \chi}}_{p_j}(t')\,\left(-\ii \vec k \cdot \c2{ (\hat{\rho\vec{u}})}_\mu(t, \vec k)\, \tilde  Z_0 [H, \tens{J}, \tens{K}]\big|_0\right)}  \nonumber \\
    & +  \sum_{\mu} \, \vec E_{\mathrm I,p_\mu}\cdot \ii\vec k^\ntop g_{qp}(0)\, \hat \rho_\mu(t, \vec k)\,\tilde  Z_0 [H, \tens{J}, \tens{K}]\big|_0 \nonumber \\
    & =  \sum_{\mu,j} \ii \int \d t'\, \vec E_{\mathrm I,p_j}\cdot \wick{\c2{\hat{\vec \chi}}_{p_j}(t')\,\left(-\ii \vec k \cdot \c2{ (\hat{\rho\vec{u}})}_\mu(t, \vec k)\right)} \, \tilde  Z_0 [H, \tens{J}, \tens{K}]\big|_0   \nonumber\\
    &+ \sum_{\mu,j}\ii \int \d t'\,  E_{\mathrm I,{\mathcal H_j}}\, \wick{\c2{\hat{\chi}}_{{\mathcal H}_j}(t')\,\left(-\ii \vec k \cdot \c2{ (\hat{\rho\vec{u}})}_\mu(t, \vec k)\right) } \, \tilde Z_0 [H, \tens{J}, \tens{K}]\big|_0 \;,\nonumber
    \end{align}
    where in the last step we used $g_{qp}(0)=0$ to cancel the second term and instead added a term that is identically zero. It vanishes since the contraction of
    $\hat{\chi}_{{\mathcal H}_j}$ with a velocity-density operator is identically zero. 
    The zero-term allows us to rewrite the expression as a single interaction operator acting on the velocity density field. Rewriting them in terms of expectation values, the density evolution equation is found to be the continuity equation independent of the exact form of the interactions
    \begin{align}\label{eff_firstorder_density}
    \partial_t \ave{\rho(t, \vec k )}^{(1)} =
    - \ii \vec k \cdot\ave{\rho\vec u (t,\vec k)}^{(1)}\;.
    \end{align}
    
    Proceeding to the next equation of motion, the time evolution of the velocity-density is
    \begin{align}
    \partial_t \ave{\rho \vec u (t, \vec k)}^{(1)}
    &= \, \partial_t\, \sum_{\mu,j}\ii \int \d t' \,\vec E_{\mathrm I,p_j} \cdot \wick{\c2{\hat{\vec \chi}}_{p_j}(t')\,\c2{\hat\rho}_\mu}(t, \vec k)\,\hat{\vec u}_\mu(t) \nonumber \\
    & + \partial_t\, \sum_{\mu,j}\ii \int \d t' \,\vec E_{\mathrm I,p_j} \cdot \wick{\c2{\hat{\vec \chi}}_{p_j}(t')\,\hat\rho_\mu(t, \vec k)\,\c2{\hat {\vec u}}_\mu(t)}\nonumber \\
    & + \partial_t\, \sum_{\mu,j}\ii \int \d t' \, E_{\mathrm I,{\mathcal H_j}} \cdot \wick{\c2{\hat{\chi}}_{{\mathcal H}_j}(t')\,\c2{\hat\rho}_\mu(t, \vec k)\,\hat{\vec u}_\mu(t)}\nonumber \\
    & + \partial_t\, \sum_{\mu,j}\ii \int \d t' \, E_{\mathrm I,{\mathcal H_j}}  \cdot \wick{\c2{\hat{\chi}}_{{\mathcal H}_j}(t')\,\hat\rho_\mu(t, \vec k)\,\c2{\hat{\vec u}}_\mu(t)}\;.
    \end{align}
    Here, the last two terms evaluate to zero because the contained contractions vanish identically. In the first two terms the time derivatives act on the field operators and their contractions. Using the time derivatives of the fields (\ref{tder_density_field}) and (\ref{tder_velocity}) and their contractions (\ref{tder_contr_p_rho}) and (\ref{tder_contr_p_u}) as well as $g_{qp}(0)= 0$, the evolution equation reads
    \begin{align}
    \partial_t &\ave{\rho \vec u (t, \vec k)}^{(1)}\\
    & = \sum_{\mu,j}\ii \int \d t'\, \vec E_{\mathrm I,p_j}\cdot \wick{\c2{\hat{\vec \chi}}_{p_j}(t')\,\left(-\ii \vec k \cdot \c2{ (\hat{\rho\vec{u}})}_\mu(t, \vec k)\right)\,\hat {\vec u}_\mu(t)  }\nonumber \\
    &+ \sum_{\mu,j}\ii \int \d t'\, \vec E_{\mathrm I,p_j} \cdot \wick{\c2{\hat{\vec \chi}}_{p_j}(t')\,\left(-\ii \vec k \cdot (\hat{\rho\vec{u}})_\mu(t, \vec k)\right)\,\c2{\hat{\vec u}}_\mu(t)} \nonumber \\
    & + \sum_{\mu} \, \vec E_{\mathrm I,p_\mu} \tfrac{-1}{m} \,g_{pp}(0) \,\hat \rho_\mu(t, \vec k) \nonumber\;.
    \end{align}
    One can now add corresponding $E_{\mathrm I_\mathcal H} \chi_{\mathcal H_j}$-terms to the first two terms in the above expression, such that the full interaction operators can be identified. This is again allowed because the additional terms contain contractions that evaluate to zero. With the complete interaction operator, the above expression can be rewritten in terms of expectation values. This yields
    \begin{align}\label{eff_firstorder_velocity}
    \partial_t \ave{\rho \vec u (t, \vec k)}^{(1)} =&
    - \ii \vec k \cdot \ave{\rho\vec u \otimes \vec u(t,\vec k)}^{(1)}
    - \ave{ \tfrac{1}{m}\vec E_{\mathrm I_p}\,  \rho(t, \vec k )} ^{(0)} \;,
    \end{align}
    where the last term containing the interacting terms of the equation of motion is reduced by one order in perturbation theory.
    
    Finally, the time evolution of the energy-density is
    \begin{align}
    \partial_t& \, \ave{\varepsilon (t, \vec k)}^{(1)}   = \,\partial_t \, \ave{\tfrac{3}{5} \mathcal H \rho(t,\vec k ) }^{(1)}  \\
    & =   \,\partial_t \, \tfrac{3}{5}\sum_{\mu,j}\ii \int \d t'\, \vec E_{\mathrm I,p_j}\cdot\wick{\c2{ \hat{\vec \chi}}_{p_j}(t')\, \c2{\hat\rho}_\mu(t, \vec k)\,\hat {\mathcal H}_\mu(t) }\nonumber\\
    &+  \partial_t \, \tfrac{3}{5}\sum_{\mu,j}\ii \int \d t'\, \vec E_{\mathrm I,p_j}\cdot \wick{\c2{\hat{\vec \chi}}_{p_j}(t')\, \hat\rho_\mu(t, \vec k)\,\c2{\hat {\mathcal H}}_\mu(t)}\nonumber \\
    &+  \partial_t \,\tfrac{3}{5} \sum_{\mu,j}\ii \int \d t'\,  E_{\mathrm I_{\mathcal H_j}} \cdot\wick{\c2{ \hat{ \chi}}_{{\mathcal H}_j}(t')\, \c2{\hat\rho}_\mu(t, \vec k)\,\hat {\mathcal H}_\mu(t)}\nonumber \\
    &+ \partial_t \, \tfrac{3}{5} \sum_{\mu,j}\ii \int \d t'\,  E_{\mathrm I,{\mathcal H_j}}\cdot \wick{\c2{\hat{\chi}}_{{\mathcal H}_j}(t')\, \hat\rho_\mu(t, \vec k)\,\c2{\hat{\mathcal H}}_\mu(t)}\;, \nonumber
    \end{align}
    where the second and third terms vanish identically. In the remaining terms the expressions for time derivatives of the field operators and contractions \eqref{tder_density_field} to \eqref{tder_contr_h_h} can be inserted. This yields
    \begin{align}
    \partial_t & \ave{\varepsilon  (t, \vec k)}^{(1)}   \\
    &= \, \tfrac{3}{5}\,\sum_{\mu,j}\ii \int \d t'\, \vec E_{\mathrm I,p_j} \cdot \wick{\c2{\hat{\vec \chi}}_{p_j}(t')\, \left(-\ii \vec k \cdot \c2{(\hat{\rho\vec{u}})}_\mu(t, \vec k)\right)\,\hat {\mathcal H}_\mu(t)} \nonumber\\
    &+ \tfrac{3}{5} \sum_{\mu,j}\ii \int \d t'\, \vec E_{\mathrm I,p_j}\cdot \wick{\c2{\hat{\vec \chi}}_{p_j}(t')\, \left(-\ii \vec k \cdot {(\hat{\rho\vec{u}})}_\mu(t, \vec k)\right)\,\c2{\hat {\mathcal H}}_\mu(t)}\nonumber \\
    &+ \tfrac{3}{5} \sum_{\mu,j}\ii \int \d t'\,  E_{\mathrm I, {\mathcal H_j}} \cdot \wick{\c2{\hat{\chi}}_{{\mathcal H}_j}(t')\, \left(-\ii \vec k \cdot {(\hat{\rho\vec{u}})}_\mu(t, \vec k)\right)\,\c2{\hat{\mathcal H}}_\mu(t)}\nonumber \\
    &+ \tfrac{3}{5} \sum_{\mu} \,   E_{\mathrm I,{\mathcal H_\mu}} \left(- g_{{\mathcal H}{\mathcal H}}(0)\right)  \hat\rho_\mu(t, \vec k) \nonumber \\
    &+\tfrac{3}{5}  \sum_{\mu,j}\ii \int \d t'\,  E_{\mathrm I,{\mathcal H_j}}\cdot \wick{\c2{\hat{\chi}}_{{\mathcal H}_j}(t')\, \left(-\ii \vec k \cdot \c2{ {(\hat{\rho\vec{u}})}}_\mu(t, \vec k)\right)\,\hat{\mathcal H}_\mu(t)} \;, \nonumber
    \end{align}
    where the last term, which evaluates to zero, was added once again to complete an interaction operator. The terms can then be brought into the convenient final form
    \begin{align}\label{eff_firstorder_energy}
    \partial_t \ave{\varepsilon (t, \vec k)}^{(1)} = - i\vec k \cdot \ave{\varepsilon \vec u (t, \vec k) }^{(1)} - \ave{ \tfrac{3}{5} E_{\mathrm I,{\mathcal H}}\rho(t,\vec k )}^{(0) }\;.
    \end{align}
    
    The three equations (\ref{eff_firstorder_density}), (\ref{eff_firstorder_velocity}) and (\ref{eff_firstorder_energy}) describe the evolution of
    the collective density, velocity-density, and energy-density fields of an ensemble of fluid particles with a general interaction of the form Eq.~\eqref{general_interaction_fluid_appendix}.
    
    It turns out that the additional interaction operators appearing in higher perturbative orders do not interfere with the above considerations. Some care must be taken to check whether the prefactors in the perturbation series work out correctly. A thorough derivation in $n$-th order perturbation theory proceeds analogously to the derivation of the BBGKY-hierarchy in \cite{2015PhRvE..91f2120V} and indeed shows that the combinatorics work out correctly.
    
    Instead of performing the full calculation here, we just state the general result
    \begin{align}
    &\partial_t \ave{\rho(t, \vec k )}^{(n)}  = - \ii \vec k \cdot\ave{\rho\vec u (t,\vec k)}^{(n)}\label{general_int_continuity_f}\\
    &\partial_t \ave{\rho \vec u (t, \vec k)}^{(n)} =   - \ii \vec k \cdot\ave{\rho\vec u \otimes \vec u(t,\vec k)}^{(n)} - \ave{ \tfrac{1}{m}\vec E_{\mathrm I,p}  \rho(t, \vec k )} ^{(n-1)} \label{general_int_euler_f}\\
    &\partial_t \ave{\varepsilon (t, \vec k)}^{(n)} = - i\vec k \cdot\ave{\varepsilon\vec u (t, \vec k) }^{(n)} - \ave{\tfrac{3}{5} E_{\mathrm I,{\mathcal H}}\rho(t,\vec k )}^{(n-1) } \label{general_int_energy_f}\;,
    \end{align}
    or in configuration space
    \begin{align}
    & \partial_t \ave{\rho(t, \vec q )}^{(n)}  + \, \partial_{\vec q} \cdot\ave{\rho\vec u (t,\vec q)}^{(n)}= 0\label{general_int_continuity_config}\\
    & \partial_t \ave{\rho \vec u (t, \vec q)}^{(n)}  +\, \partial_{\vec q}\cdot \ave{\rho\vec u \otimes \vec u(t,\vec q)}^{(n)}
    = - \ave{  \tfrac{1}{m}\vec E_{\mathrm I,p}  \rho(t, \vec q )} ^{(n-1)} \label{general_int_euler_config} \\
    & \partial_t \ave{\varepsilon(t, \vec q)}^{(n)}  + \,\partial_{\vec q} \cdot\ave{\varepsilon\vec u (t, \vec q) }^{(n)}
    =  - \ave{ \tfrac{3}{5} E_{\mathrm I,{\mathcal H}}\rho(t,\vec q )}^{(n-1) }\label{general_int_energy_config}\;.
    \end{align}

\section{Heuristic derivation of the interaction operator}
    \label{appendix_heuristic_derivation}
    In Section \ref{subsection_SPH_interactions} we already derived interactions mimicking fluid behaviour in an approach similar to smoothed particle hydrodynamics. Here we present a second derivation of the same interactions based on a heuristic understanding of the underlying physical processes. These interactions will again include accelerations due to pressure gradients, pressure-volume work, and diffusive effects. 
    
    \subsection{Acceleration by pressure gradients}
    To derive an interaction that describes accelerations caused by pressure gradients, we start from the collective pressure field at an arbitrary point $\vec q_2$ in the fluid. Using the relation between the enthalpy-density and the pressure in absence of any internal degrees of freedom, $P = \tfrac{2}{5} h$, and summing over all one-particle contributions, we get
    \begin{align}
    P(\vec q_2) = \sum_i \tfrac{2}{5}\, \mathcal H_i \, \rho_i(\vec q_2)\;.
    \end{align}
    At another position in the fluid $\vec q_1$ the fluid will experience the acceleration caused by the surrounding pressure field. To derive the acceleration, we imagine the surroundings of $\vec q_1$ to be divided into spherical shells. For one specific shell with radius $r_2 =|\vec q_2 -\vec q_1|$, the pressure field exerts a force on a surface element,
    \begin{align}
    \vec F(\vec q_2)_{\text{one sphere}} = -P(\vec q_2)\, \vec{\d A}_{12}\;.
    \end{align}
    Here, $\vec{\d A}_{12}$ is the outer unit surface element at position $\vec q_2$ on the spherical shell around $\vec q_1$, and the minus sign ensures that the force is directed towards the centre of the sphere. 
    
    An integral over the complete spherical surface gives the force contribution due to that surface. A radial integration finally yields the net force acting on the fluid at position $\vec q_1$, where the strength of the contribution of each sphere is described by the radial function $W(|\vec q_2-\vec q_1|)$,
    \begin{align}
    \vec F_\text{\tiny net}(\vec q_1)&= -  \int d r_2 \int \vec{\d A}_{12} \,P(\vec q_2) \,W(|\vec q_2-\vec q_1|)\nonumber \\
    &= -  \sum_i \int d r_2  \int \vec{\d A}_{12} \,\tfrac{2}{5}\, \mathcal H_i \, \rho_i(\vec q_2) \,W(|\vec q_2-\vec q_1|)\nonumber \\
    &=  \sum_i \int d^3 q_2  \,\tfrac{2}{5}\, \mathcal H_i \, \rho_i(\vec q_2) \,W(|\vec q_2-\vec q_1|) \,\hat{\vec e}_{21}\;,
    \end{align}
    where $\hat{\vec e}_{21}$ is the unit vector pointing from $\vec q_2$ to $\vec q_1$. If we now assume a spherically symmetric scalar function
    $V(\vec q_2 - \vec q_1)$, we can simplify the above expression by defining $V$ such that
    \begin{align} \label{spherical_function}
    W(|\vec q_2-\vec q_1|) \,\hat{\vec e}_{21} =: \partial_{q_2} V(\vec q_2 - \vec q_1)\;.
    \end{align}
    We then get
    \begin{align}
    \vec F_\text{\tiny net}(\vec q_1)&=  \sum_i \int d^3 q_2  \,\tfrac{2}{5}\, \mathcal H_i \, \rho_i(\vec q_2) \,\partial_{q_2} V(\vec q_2 - \vec q_1)\;.
    \end{align}
    
    So far, the acceleration is assigned to a point in the macroscopic fluid. However, at this point, we may assign the acceleration to the mesoscopic particles which make up the fluid. For particles of equal mass, the contribution of the $j$-th particle at $\vec q_1$ is described by its density at that point, $\rho_j(\vec q_1)$. The force and the associated momentum change experienced by the $j$-th particle then is
    \begin{align}\label{heurisitic_momentum_change_ideal_hydro}
    \dot{\vec p}_j= \sum_{(i,j)}\, \int d^3 q_1\, d^3 q_2  \,\tfrac{2}{5}\, \mathcal H_i \, \rho_i(\vec q_2) \, \rho_j(\vec q_1) \,\partial_{q_2}V(\vec q_2 - \vec q_1)\;,
    \end{align}
    where $(i,j)$ denotes the sum over $i$ and $j$ with $i\neq j$ to avoid self-interactions.
    

    \subsection{Pressure-volume work}
    In a similar way, pressure-volume work is modelled. For a start, we assume to be in the rest frame of the $j$-th particle. In this frame the collective velocity-density field at a position $\vec q_2$ is
    \begin{align}
    \rho \vec u (\vec q_2) = \sum_i \, \vec u_i\, \rho_i(\vec q_2)\;.
    \end{align}
    Just as before, the neighbourhood of $\vec q_1$ is divided into spherical shells. If the velocity-density field converges or diverges over the surface of each shell, they compress or expand the sphere against or with the pressure inside the sphere, respectively. This changes the local energy-density due to pressure-volume work.
    
    The pressure within the sphere can approximately be described by the pressure field at the central point $\vec q_1$ which is
    \begin{align}
    P(\vec q_1) = \sum_j \tfrac{2}{5} \mathcal H_j \,\rho_j(\vec q_1)\;.
    \end{align}
    
    The convergence or divergence of the velocity field over the surface of a spherical shell is found by a projection of the velocity-density field $\rho \vec u = \sum_i \rho_i \vec u_i$ onto the negative surface element and an integration over the entire surface. The sign is chosen such that the convergence is positive and the divergence negative. 
    
    Multiplying the pressure inside the shells with their convergence/divergence across their surface yields the net change of internal energy-density inside each shell. Another integration over the radius of the spherical shells weighted by the radial function $ \,W(|\vec q_2-\vec q_1|)$ then yields the change of internal energy at $\vec q_1$,
    \begin{align}
    \dot \varepsilon(\vec q_1)& = -  \sum_{i,j}\, \int d r_{2} \int \vec{\d A}_{12}  \cdot \vec u_i\, \rho_i(\vec q_2) P_j(\vec q_1)\,  W(|\vec q_2-\vec q_1|) \nonumber \\
    =&-  \sum_{i,k} \, \int  d r_{2} \int  \vec{\d A}_{12}\cdot \vec u_i\, \rho_i(\vec q_2) \tfrac{2}{5} \mathcal H_j\,\rho_j(\vec q_1) \,W(|\vec q_2-\vec q_1|)\nonumber\\
    = &\sum_{i,k} \, \int  d^3 q_2 \, \vec u_i\, \rho_i(\vec q_2) \tfrac{2}{5} \mathcal H_j \,\rho_j(\vec q_1)\,\partial_{q_2} \,V(\vec q_2-\vec q_1)\;.
    \end{align}
    
    The energy-density change of the fluid must now be assigned to the enthalpy change of particles. For the momentum change, we performed this step by weighting the momentum change of the fluid at a position $\vec q_1$ with the density contribution of a particle at that position and integrating over $q_1$. 
    
    In the case of the energy-density at $\vec q_1$, the above expression is already related to the pressure contribution of a particle at that position. Hence, an integration over $\mathrm d^3q_1$ serves to switch from the field description to a particle description. At the same time, the integration turns the energy-density into an enthalpy if a factor of $\frac{5}{2}$ is included. 

    Finally, one can switch from the rest frame of the $j$-th particle into a general frame by replacing all velocities by their difference to the $j$-th particle's velocity. The enthalpy change of the $j$-th particle due to pressure-volume work becomes 
    \begin{align}\label{heuristic_enthalpy_change_ideal_hydro}
    \dot{\mathcal H}_j  = \sum_{(i, j)}\int d^3 q_1 \, d^3 q_2 \, (\vec u_i- \vec u_j) \, \rho_i(\vec q_2) \tfrac{2}{3} \mathcal H_j \,\rho_j(\vec q_1)\,\partial_{q_2} \,V(\vec q_2-\vec q_1)\;.
    \end{align}

    \subsubsection{Diffusive currents}
    
    Diffusive currents appear if gradients in the energy and momentum fields exist. The diffusion of microscopic particles leads to a net transport of these quantities. Here, we first derive these currents and in a second step model their impact on the dynamics of the mesoscopic particles.
    
    We start from the collective energy and momentum-density fields at position $\vec q_2$. They are found by a summation over all one-particle contributions at that position
    \begin{align}
    & \varepsilon (\vec q_2) = \sum_i\, \tfrac{3}{5} \mathcal H_i \,\rho_i(\vec q_2)\;,\\
    &\vec p ( \vec q_2) = \sum_i\, \vec p_i \,\rho_i(\vec q_2)\;.
    \end{align}
    Note that $\vec p(\vec q_2)$ can be interpreted as three independent fields, one for each component, being the source of separate diffusive currents.
    
    To calculate the diffusive current at position $\vec q_3$, we once again decompose its surroundings into spherical shells. For each of these shells, we can approximate the gradient of the field by a surface integral weighted by the volume of the sphere (in the limit of an infinitely small shell, i.e.\ a vanishing volume, this would be exact),
    \begin{align}
    &\partial_{\vec q_3} \,\varepsilon (\vec q_3) \approx \sum_i\, \underset{\partial V_{32}}{\int}\vec{\d A}_{32} \tfrac{3}{5} \mathcal H_i \,\rho_i(\vec q_2)\;,\\
    &\partial_{q_3}^a \, p^b ( \vec q_3) \approx \sum_i\,\underset{\partial V_{32}}{\int} \d A^a_{32}\, p_i^b \,\rho_i(\vec q_2)\;.
    \end{align}
    Here,  $\vec{\d A}_{32}$ is the outer unit surface element and $V_{32}$ the volume of the sphere around $\vec q_3$ with $\vec q_2$ on its surface. The indices $a$ and $b$ denote the vector components.
    
    An integration over all spheres weighted by a radial kernel $W_\text{\tiny D}$ yields the diffusive energy current $\vec j$ and the three momentum currents that are combined into a diffusive stress-energy tensor $T^{ab}$. They read
    \begin{align}
    &\vec j (\vec q_3) =-\tau \sum_i \int \d r_{32}\,  \underset{\partial V_{32}}{\int}\vec{\d A}_{32} \tfrac{3}{5} \mathcal H_i \,\rho_i(\vec q_2)\, W_\text{\tiny D}(r_{32}) \;,\\
    &T^{ab} (\vec q_3) = - c_{ab} \sum_i \int \d r_{32} \, \underset{\partial V_{32}}{\int} \left( \d A_{32}^a \, p_i^b\right)_\text{\tiny s} \, \rho_i(\vec q_2)\, W_\text{\tiny D}(r_{32})\;,
    \end{align}
    where $\tau$ and $c_{ab}$ are diffusion constants.
    
    We introduce a spherically symmetric scalar function $V_\text{\tiny D}(\vec q_2 - \vec q_3)$ such that $\partial_{\vec q_3}V_\text{\tiny D}(\vec q_2 - \vec q_3)  = W_\text{\tiny D}(\vec q_2 - \vec q_3) $ and hence
    \begin{align}
    &\vec j (\vec q_3) =\tau \sum_i \int \d^3 q_2\,  \tfrac{3}{5} \mathcal H_i \,\rho_i(\vec q_2)\, \partial_{q_2}V_\text{\tiny D}(\vec q_2 - \vec q_3) \;,\\
    & T^{ab} (\vec q_3) =  c_{ab} \sum_i \int \d^3 q_2 \, \rho_i(\vec q_2) \, \left( \partial_{ q_2}^a \, p_i^b\right)_\text{\tiny s} V_\text{\tiny D}(\vec q_2 - \vec q_3)\;.
    \end{align}

    To describe the diffusive currents in the particle-based approach, they must once again be associated with the particles at position $\vec q_3$. For the $k$-th particle we get
    \begin{align}
    \label{heuristic_diffusive_energy_current}
    \vec j_k  &= \int \d^3 q_3 \, \vec j (\vec q_3) \, \rho_j(\vec q_3)  \\
    &=\tau \sum_i \int \d^3 q_2\,  \d^3 q_3 \, \tfrac{3}{5} \mathcal H_i \,\rho_i(\vec q_2)\, \partial_{q_2}V_\text{\tiny D}(\vec q_2 - \vec q_3) \, \rho_k (\vec q_3)\;,\nonumber\\
    \label{heuristic_diffusive_momentum_currents}
    T^{ab}_k &= \int \d^3 q_3 \, T^{ab} (\vec q_3) \, \rho_j(\vec q_3)\\
    & =  c_{ab} \sum_i \int \d^3 q_2\, \d^3 q_3  \, \rho_i(\vec q_2) \, \left( \partial_{ q_2}^a \, p_i^b\right)_\text{\tiny s} V_\text{\tiny D}(\vec q_2 - \vec q_3)\, \rho_k (\vec q_3)\;. \nonumber
    \end{align}

    \subsubsection{Dynamical impact on the mesoscopic particles}

    The diffusive currents influence the surrounding mesoscopic particles. To quantify this effect, we again decompose the surroundings of a point $\vec q_1$ in the fluid into spherical shells. The fluid at this point is accelerated/its energy changes if there is a net inflow of the diffusive momentum/energy currents through its surface. For each shell, the inflow can be calculated by projecting the currents onto the negative outer surface element and integrating over the whole sphere. An integration over all spheres weighted by a radial kernel then yields the internal energy and momentum changes of the fluid at position $\vec q_1$
    \begin{align}
    &\dot p^a (\vec q_1) = -\sum_k \int \d r_{14}\underset{\partial V_{14}}{\int} \d A_{14}^b \, T^{ab}_k \, \rho_k(\vec q_4)\, W(|\vec q_4 - \vec q_1|) \,, \\
    &\dot {\mathcal E}(\vec q_1) = -\sum_k \int \d r_{14}\underset{\partial V_{14}}{\int} \vec{\d A}_{14}\cdot  \vec j_k \, \rho_k(\vec q_4) \, W(|\vec q_4 - \vec q_1|)\;.
    \end{align}
    We can again write
    \begin{align}
    &\dot p^a (\vec q_1) =  \sum_ k \int \d ^4 q_4  \,T^{ab}_k\, \rho_k (\vec q_4)\, \partial_{q_4}^b V(\vec q_4 - \vec q_1)\,,\\
    &\dot {\mathcal E}(\vec q_1) = \sum_ k \int \d ^3 q_4  \,\vec j_k \, \rho_k (\vec q_4) \cdot \partial_{\vec q_4} V(\vec q_4 - \vec q_1)\;.
    \end{align}
    
    These changes can be associated with a particle by weighting each position in the fluid with the contribution of the particle to the density at that point and integrating over the entire space. For the momentum and enthalpy change of the $j$-th particle this means
    \begin{align}\label{heurisitic_viscous_current_momentum_change}
    \dot{p}^a_j=& \int \d^3 q_1 \dot p^a(\vec q_1)\, \rho_j(\vec q_1) \\
    =& c_{ab}  \sum_ {(i,k)} \int \d^3 q_1  \,  \d^3 q_2\, \d^3 q_3  \, \d ^4 q_4 \, \left( \partial_{ q_2}^a \, p_i^b\right)_\text{\tiny s} V_\text{\tiny D}(\vec q_2 - \vec q_3)\nonumber\\
    & \times\, \partial_{q_4}^b V(\vec q_4 - \vec q_1)\, \rho_j(\vec q_1) \, \rho_i(\vec q_2) \,  \rho_k (\vec q_3) \, \rho_k (\vec q_4)\, \nonumber
    \end{align}
    
    and
    \begin{align}
    \dot{\mathcal H}_j=& \tfrac{5}{3} \int \d^3 q_1 \dot{\mathcal E}(\vec q_1)\, \rho_j(\vec q_1) \\
    =&   \tau \sum_{(i,k)} \int \d^3 q_1\,   \d^3 q_2\,  \d^3 q_3 \, \d ^3 q_4  \mathcal H_i \, \partial_{q_2}V_\text{\tiny D}(\vec q_2 - \vec q_3) \nonumber \\
    & \times  \partial_{\vec q_4} V(\vec q_4 - \vec q_1)\, \rho_j(\vec q_1) \,\rho_i(\vec q_2)\, \rho_k (\vec q_3) \, \rho_k (\vec q_4) \nonumber\;.
    \end{align}
    
    Apart from these two effects, we need to include a third impact caused by the diffusive currents. The stress-energy-tensor has the units of a momentum current density, i.e.\ the units of a pressure. Hence, the diffusive momentum currents act as an effective pressure that causes pressure-volume work. This process extracts energy from the macroscopic currents and turns it into internal energy. In other words, it describes the internal energy rise due to diffusive friction.
    
    From the interactions of ideal hydrodynamics, we already know an interaction that describes pressure-volume work, namely Eq.~\eqref{heuristic_enthalpy_change_ideal_hydro}. To model viscous pressure-volume the scalar pressure field $P_j(\vec q_1)= \tfrac{2}{5} \mathcal H_j \rho_j(\vec q_1)$ in that term must be replaced by the diffusive stress energy tensor,
    \begin{align}\label{heuristisch_pressure_volume_work_viscous}
    \dot{\mathcal H}_j  = &\tfrac{5}{3} \sum_{i}\int \d^3 q_1 \, \d^3 q_2 \, ( u_i^a- u_j^a) \, T^{ab}_j \,\\
    &\times \partial_{q_2}^b \,V(\vec q_2-\vec q_1)\, \rho_i(\vec q_2)\,\rho_j(\vec q_1)\nonumber\\
    = &\tfrac{5}{3} c_{ab}\sum_{(i,l)}\int \d^3 q_1 \, \d^3 q_2  \, \d^3 q_3  \, \d^3 q_4\, \left( \partial_{ q_3}^a \, p_l^b\right)_\text{\tiny s} V_\text{\tiny D}(\vec q_3 - \vec q_4)\nonumber \\
    &\times  \,\partial_{q_2}^b \,V(\vec q_2-\vec q_1)\,  ( u_i^a- u_j^a) \,\rho_l(\vec q_3) \, \rho_i(\vec q_2)   \, \rho_j(\vec q_4)  \,\rho_j(\vec q_1) \nonumber\;.
    \end{align}
    At this point one might wonder why we used the formal equivalence between the pressure and the diffusive stress energy tensor only in the case of pressure-volume work and not for accelerations caused by pressure gradients. The latter, however, would have led to exactly the same interaction as we found in Eq.~\eqref{heurisitic_viscous_current_momentum_change}, only with a slightly different argumentation in its derivation.
    
    In the last step, both the momentum and enthalpy change of the $j$-th particle can be simplified by using the relation between two Dirac distributions \eqref{delta_relation}. 
    The momentum change of the $j$-th particle then is
    \begin{align} \label{heuristic_momentum_change_viscous}
    \dot{p}^a_j=& c_{ab}  \sum_ {(i, k)} \int \d^3 q_1  \,  \d^3 q_2\, \d^3 q_3  \, \left( \partial_{ q_2}^a \, p_i^b\right)_\text{\tiny s} V_\text{\tiny D}(\vec q_2 - \vec q_3)\\
    & \times\, \partial_{q_3}^b V(\vec q_3 - \vec q_1)\, \rho_j(\vec q_1) \, \rho_i(\vec q_2) \,  \rho_k (\vec q_3) \;,\nonumber
    \end{align}
    and the enthalpy change (combining both of the above effects) is
    \begin{align}            \label{heuristic_enthalpy_change_viscous}
    \dot{\mathcal H}_j & = \tau \sum_{(i, k)} \int \d^3 q_1\,   \d^3 q_2\,  \d^3 q_3 \, \mathcal H_i \, \partial_{q_2}V_\text{\tiny D}(\vec q_2 - \vec q_3)\\
    & \times  \partial_{\vec q_3} V(\vec q_3 - \vec q_1)\, \rho_j(\vec q_1) \,\rho_i(\vec q_2)\, \rho_k (\vec q_3) \nonumber \\
    &+ \tfrac{5}{3} c_{ab}\sum_{(i,l)}\int \d^3 q_1 \, \d^3 q_2  \, \d^3 q_3 \, \left( \partial_{ q_3}^a \, p_l^b\right)_\text{\tiny s} V_\text{\tiny D}(\vec q_3 - \vec q_1)\nonumber \\
    &\times  \,\partial_{q_2}^b \,V(\vec q_2-\vec q_1)\,  ( u_i^a- u_j^a) \,\rho_l(\vec q_3) \, \rho_i(\vec q_2)  \,\rho_j(\vec q_1) \;. \nonumber
    \end{align}
    \vspace{1em}
    
    \subsection{Comparison of the two derivations}
    Eq.~\eqref{SPH_momentum_change_ideal_hydro} and Eq.~\eqref{SPH_enthalpy_change_ideal_hydro} describe the momentum and enthalpy change of the $j$-th particle in ideal hydrodynamics derived in the SPH-like approach while Eq.~\eqref{heurisitic_momentum_change_ideal_hydro} and Eq.~\eqref{heuristic_enthalpy_change_ideal_hydro} describe the same quantities in the heuristic derivation.
    
    These interactions look similar in the two derivations but do not match completely which is due to the interpretation of the functions $v$ in the SPH derivation and $V$ in the heuristic one. The former was introduced as a three-dimensional density kernel and has the dimension of $[v] = \text{m}^{-3}$, the latter was constructed from the radial dependence of the interaction between two particles and is dimensionless.
    For
    \begin{align}
    \partial_{q_1} \, V(\vec q_1 - \vec q_2) = \frac{1}{\rho(\vec q_1)}\, \partial_{q_1} \, v(\vec q_1 - \vec q_2)\;,
    \end{align}
    the interactions agree. The different particle index at the enthalpy is caused by choosing an alternative discretization for an derivative acting on a velocity field,
    \begin{align}\label{velocity_discretisation_2}
    \dot{\mathcal H } (\vec q_1) &= - \frac{5}{3} \, \frac{P(\vec q_1)}{\rho(\vec q_1)}\, \left[\partial_{q_1}  \frac{\rho \vec u (\vec q_1)}{\rho(\vec q_1)}\right] \\
    &= - \frac{5}{3} \, \frac{P(\vec q_1)}{\rho^2(\vec q_1)}\, \left[\partial_{q_1}  \cdot \rho \vec u (\vec q_1) - \vec u (\vec q_1) \cdot  \partial_{q_1}  \, \rho(\vec q_1) \right] \nonumber\;.
    \end{align}
    With the equivalent choice Eq.~\eqref{velocity_discretisation_1} the index also agrees. 
    
    A similar correspondence can be found for the interactions modeling viscous effects. These are Eq.~\eqref{SPH_momentum_change_viscous} and Eq.~\eqref{SPH_enthalpy_change_viscous} for the viscous momentum change and viscous enthalpy change in the SPH derivation, and Eq.~\eqref{heuristic_momentum_change_viscous} and Eq.~\eqref{heuristic_enthalpy_change_viscous} for the same quantities in the heuristic derivation. The differences, especially that a three-density-term appears is again caused by the exact choice of discretization. 
    
    The terms agree up to a partial integration if one uses the above relation between $v$ and $V$, as well as $V_\text{\tiny D}(\vec q - \vec q') = \dirac(\vec q - \vec q')$ for the momentum change and first term of the enthalpy change, i.e.\ $V_\text{\tiny D}(\vec q - \vec q') = V(\vec q - \vec q')$. 
    

	\bibliography{literature}

\end{document}